\documentclass[sigconf, nonacm]{acmart}

\usepackage[utf8]{inputenc} 
\usepackage[T1]{fontenc}    
\usepackage{url}            
\usepackage{booktabs}       
\usepackage{nicefrac}       
\usepackage{microtype}      
\usepackage{xcolor}         

\usepackage{xspace}
\usepackage{hyperref}
\usepackage{amsmath,amsfonts}

\usepackage{censor}
\usepackage{textcomp}
\usepackage{nicefrac}       
\usepackage{caption}

\usepackage{graphicx}
\usepackage{blindtext}
\usepackage[inkscapelatex=false]{svg}
\usepackage{verbatim}
\usepackage{color,soul}
\usepackage{graphics}
\usepackage{listings}
\usepackage{multirow}
\usepackage{tabularx}
\usepackage[toc,page]{appendix}
\usepackage{mwe}%
\usepackage{todonotes}
\usepackage{tikz}
\usetikzlibrary{arrows.meta, positioning, shapes.multipart, fit, calc}
\usepackage{algorithm}
\usepackage{algpseudocode}
\usepackage{pgfplots}
\usepackage{siunitx}
\pgfplotsset{compat=newest}
\usepgfplotslibrary{groupplots}
\definecolor{darkred}{rgb}{0.55,0.0,0.0}
\pgfplotsset{compat=1.18}  
\usepackage{breakurl} 
\usepackage{xurl}

\newcommand\vldbdoi{XX.XX/XXX.XX}
\newcommand\vldbpages{XXX-XXX}
\newcommand\vldbvolume{14}
\newcommand\vldbissue{1}
\newcommand\vldbyear{2020}
\newcommand\vldbauthors{\authors}
\newcommand\vldbtitle{\shorttitle} 
\newcommand\vldbavailabilityurl{https://github.com/arupcsedu/AAFLOW}
\newcommand\vldbpagestyle{plain} 

\begin{document}
\title{[AAFLOW+] Stateful Operator Abstraction with Zero-Copy Distributed KV Cache Orchestration for Multi-Agent Workflows}

\author{Arup Kumar Sarker}
\orcid{0009-0006-2807-4986}
\author{Alexander James Halpern}
\author{Mills Staylor}

\affiliation{%
  \institution{University of Virginia, \\ Biocomplexity Institute and Initiative}
  \city{Charlottesville}
  \state{VA}
  \postcode{22904}
  \country{USA}
}
\email{djy8hg@virginia.edu} 
\email{halperna22@gmail.com}
\email{qad5gv@virginia.edu}

\author{Gregor von Laszewski}
\author{Geoffrey Fox}
\author{Yue Cheng}
\affiliation{%
  \institution{Biocomplexity Institute and Initiative \\ University of Virginia}
  \city{Charlottesville}
  \state{VA}
  \postcode{22904}
  \country{USA}
}
\email{laszewski@gmail.com}
\email{vxj6mb@virginia.edu}
\email{mrz7dp@virginia.edu}

\author{Aymen Alsaadi}
\author{Shantenu Jha}
\affiliation{%
  \institution{Rutgers University}
  \institution{Princeton Plasma Physics Laboratory}
  \city{Princeton}
  \state{NJ}
  \country{USA}
}
\email{aymen.alsaadi@rutgers.edu}
\email{shantenu.jha@rutgers.edu}

\begin{abstract}
Multi-agent LLM systems increasingly integrate retrieval, planning, and reasoning, but remain fundamentally text-centric, requiring agents to repeatedly recompute shared context through expensive prefill. Although single-request inference is known to be accelerated by KV-cache management, it is usually restricted to local serving scopes. We introduce AAFLOW+, a stateful extension of agentic workflow operators that makes KV cache a first-class distributed systems object. AAFLOW+ builds processes into communication-aware graphs that concurrently optimize data, prompts, and reusable model state. It also provides operators for KV materialization, transfer, fork, composition, and eviction. Its runtime enables zero-copy, transfer-aware execution, allowing agents to reuse long context without recomputation. AAFLOW+ reduces TTFT by up to 50.2$\times$, achieves up to 7.63$\times$ reduced multi-agent compute cost at 16-agent scale, reduces KV memory by 1.72–6.10$\times$, and increases throughput by more than 7.74$\times$, based on an analytical cost model parameterized by empirical hardware microbenchmarks. The results demonstrate that KV transmission outperforms recomputation on networks with moderate to high bandwidth, making sure KV-state sharing greatly increases efficiency in multi-agent LLM systems by replacing text passing.
\end{abstract}

\maketitle

\pagestyle{\vldbpagestyle}
\begingroup\small\noindent\raggedright\textbf{PVLDB Reference Format:}\\
\vldbauthors. \vldbtitle. PVLDB, \vldbvolume(\vldbissue): \vldbpages, \vldbyear.\\
\href{https://doi.org/\vldbdoi}{doi:\vldbdoi}
\endgroup
\begingroup
\renewcommand\thefootnote{}\footnote{\noindent
This work is licensed under the Creative Commons BY-NC-ND 4.0 International License. Visit \url{https://creativecommons.org/licenses/by-nc-nd/4.0/} to view a copy of this license. For any use beyond those covered by this license, obtain permission by emailing \href{mailto:info@vldb.org}{info@vldb.org}. Copyright is held by the owner/author(s). Publication rights licensed to the VLDB Endowment. \\
\raggedright Proceedings of the VLDB Endowment, Vol. \vldbvolume, No. \vldbissue\ %
ISSN 2150-8097. \\
\href{https://doi.org/\vldbdoi}{doi:\vldbdoi} \\
}\addtocounter{footnote}{-1}\endgroup

\ifdefempty{\vldbavailabilityurl}{}{
\vspace{.3cm}
\begingroup\small\noindent\raggedright\textbf{PVLDB Artifact Availability:}\\
The source code, data, and/or other artifacts have been made available in \texttt{AAFLOW/stateful\_agentic\_algebra} directory at \url{\vldbavailabilityurl}.
\endgroup
}


\section{Introduction}
\label{sec:introduction}
Large language models are increasingly deployed as \emph{agentic systems} that interleave retrieval, reasoning, tool invocation, and memory across multiple stages. Frameworks such as ReAct, Reflexion, AutoGen, and DSPy expand this design space, while RAG improves factual grounding by conditioning generation on external data~\cite{lewis2020rag, yao2023react, shinn2023reflexion, wu2023autogen, khattab2023dspy}. However, as these workflows become deeper and more collaborative, a fundamental systems bottleneck emerges: agents continue to communicate primarily through \emph{text}. Even when shared context has already been computed, downstream agents must replay it through costly prefill computation, repeatedly reconstructing identical model state.

This inefficiency is closely tied to the treatment of the key-value (KV) cache, a central object in LLM inference and a key component of AI memory~\cite{aimemory0526}. Recent serving systems such as vLLM~\cite{kwon2023vllm}, SGLang~\cite{zheng2023sglang}, DistServe, MemServe, Mooncake, ChunkAttention, KVCOMM, and RelayCaching ~\cite{zhong2024distserve,hu2024memserve,qin2024mooncake,ye2024chunkattention, kvcomm25ye, geng2026relaycaching} demonstrate that KV cache management is critical for reducing latency and improving throughput. These systems optimize KV reuse through techniques such as block-level allocation, prefix sharing, and decoupled prefill and decode stages. However, their scope is largely confined to single-request or single-cluster execution. They do not provide a workflow-level abstraction for explicitly transferring, branching, and reusing KV state across multiple agents, which impacts pipeline execution. In planner–retriever–solver workflows, multiple agents often operate over the same long context prefix. In tree-of-thought or debate settings, agents branch from shared prefixes and explore alternative reasoning paths. In collaborative RAG, agents independently retrieve overlapping evidence and reprocess similar context. In all cases, current systems serialize shared state back into text, transforming a stateful execution problem into repeated prompt replay. The result is increased time-to-first-token (TTFT), duplicated computation, higher framework overhead, and limited control over state placement and reuse.

\begin{figure}[htpb]
    \begin{center}
    \includegraphics[width=0.85\linewidth]{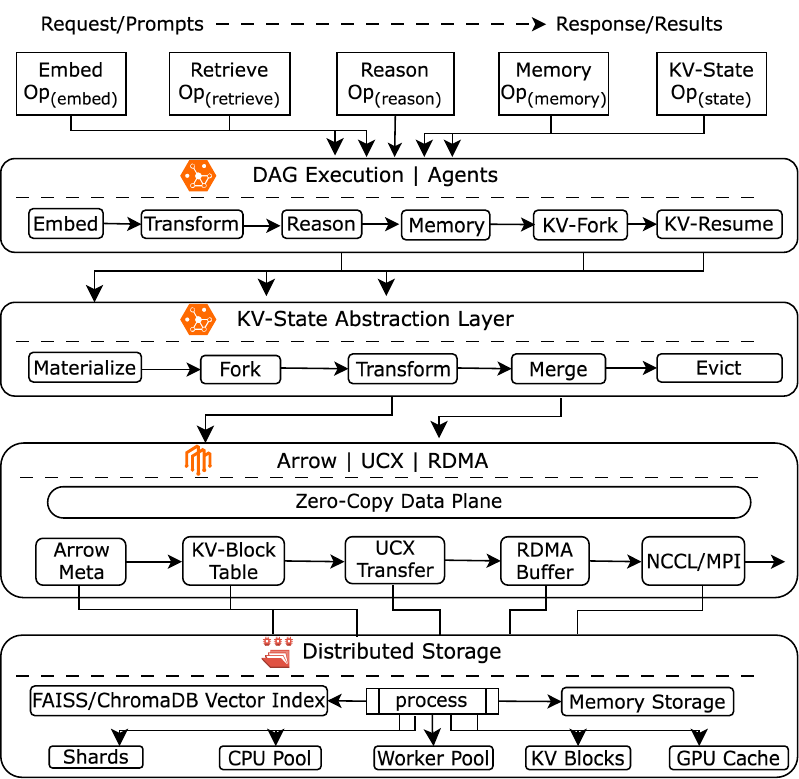}
    \caption{Layered architecture of stateful operator abstraction. The design extends AAFLOW-style operator execution with explicit KV-state materialization, fork, transfer, merge, and eviction.} 
    \label{fig:stateful_layered_architecture}
    \end{center}
\end{figure}

Recent frameworks like AAFLOW~\cite{sarker2026aaflow}, model agentic workflows as compositions of distributed operators and reduce orchestration overhead via communication-aware DAG execution and zero-copy data transfer. However, AAFLOW remains data-centric and does not expose internal model state, such as KV cache, as part of its abstraction. We introduce \textbf{AAFLOW+}, a stateful extension that elevates KV cache to a first-class distributed systems object. AAFLOW+ extends operator abstraction from \emph{dataflow} to \emph{stateflow}, enabling explicit modeling of KV-state lifecycle through operators for materialization, transfer, fork, restricted composition, and eviction (Figure~\ref{fig:stateful_layered_architecture}). This lets the compiler choose between text passing, KV reuse, state transfer, and eviction, shifting optimization from prompt construction to explicit state management. Our runtime makes the KV cache transportable by using explicit metadata, zero-copy communication, and transfer-aware scheduling. By separating metadata from tensor buffers and using Arrow-based formats, the system minimizes serialization costs and enables efficient state transfer between distributed components~\cite{apachearrow}. AAFLOW+ ensures correctness by enforcing compatibility on model identity, positional encoding, and execution history, allowing safe prefix reuse and controlled state composition.

We evaluate AAFLOW+ on GPU clusters using Mistral-7B and Llama-3-8B. Results show that stateful execution significantly improves performance over text-based orchestration. AAFLOW+ reduces TTFT by up to \textbf{50.2$\times$} at long context, achieves up to \textbf{7.63$\times$} lower multi-agent latency at 16-agent scale, reduces peak KV memory by \textbf{1.72$\times$--6.10$\times$}, and improves throughput by over \textbf{7.74$\times$}. Transfer–recompute analysis shows that KV transfer dominates recomputation on moderate-to-high bandwidth networks, while consistency experiments preserve deterministic output agreement. These results show that workflow-level KV-state sharing reduces latency, memory pressure, and framework overhead while complementing existing LLM serving systems.
\section{Background and Motivation}

\subsection{From Agentic Dataflow to Stateful Execution}

Agentic large language model (LLM) applications are increasingly organized as multi-stage workflows in which retrieval, reasoning, tool use, memory access, and response generation interact dynamically. While frameworks like DSPy treat language-model applications as compositional programs that can be optimized through compilation~\cite{wu2023autogen,khattab2023dspy}, systems like AutoGen and LangGraph offer flexible multi-agent programming abstractions. Nonetheless, the majority of agent frameworks continue to be largely text-centric, with agents exchanging serialized strings representing tool outputs, recovered excerpts, intermediate summaries, and natural language messages. Although this approach is practical at the application layer, it conceals a more serious inefficiency in the system. A downstream agent typically needs to re-tokenize and re-prefill the same context before it can produce fresh output when it receives text that has previously been handled by an upstream model invocation.

Instead of viewing agentic workflows as loosely connected framework callbacks, AAFLOW~\cite{sarker2026aaflow} treated them as compilable distributed programs. In AAFLOW, a workflow is represented as a set of operators,
\begin{equation}
W = \{Op_{embed}, Op_{retrieve}, Op_{reason}, Op_{memory}, Op_{upsert}\},
\end{equation}
where each operator is defined as
\begin{equation}
Op_i = (I_i, O_i, f_i, P_i).
\end{equation}
Here, $I_i$ and $O_i$ denote the input and output objects, $f_i$ is the transformation function, and $P_i$ is the communication pattern associated with the operator. This abstraction allows the runtime to lower an agentic workflow into an execution graph,
\begin{equation}
G = Compile(W),
\end{equation}
where communication patterns such as broadcast, shuffle, reduce, and embarrassingly parallel execution become explicit parts of the execution plan. The key insight of this paper is that the same principle should apply not only to external workflow data, such as documents, embeddings, and vector indices, but also to \textit{internal model execution state}. The state created during LLM prefill is typically locked inside a local serving runtime in current agentic systems. Because of this, multi-agent workflows maximize the flow of data throughout the model, but they are unable to optimize the transportation and reuse of the key-value (KV) cache, the model's most costly intermediate artifact.

\begin{figure}[htpb]
    \begin{center}
    \includegraphics[width=0.85\linewidth]{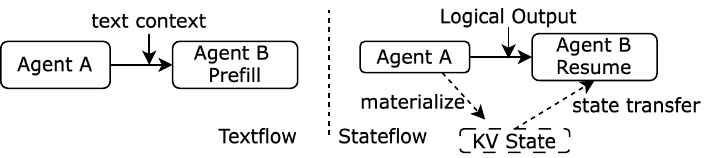}
    \caption{Textflow forces downstream agents to replay context, whereas stateflow transfers reusable KV execution state.} 
    \label{fig:text_vs_stateflow}
    \end{center}
\end{figure}

\subsection{Cost of Text-Based Agent Communication}

Consider a two-agent workflow where an upstream agent gets documents and creates a lengthy context, and a downstream agent uses that context to perform reasoning in order to comprehend the limitations of text-based communication. In existing systems, text is usually emitted by the upstream agent and consumed as a new prompt by the downstream agent. Before generating its first output token, the downstream model must make a prefill pass over the whole context if the prompt length is $L$. We denote this cost as
\begin{equation}
T_{prefill}(L).
\end{equation}
The total execution time for a single agent invocation can therefore be approximated as
\begin{equation}
T_{agent} = T_{prefill}(L) + T_{decode}(Y) + \Omega,
\end{equation}
where $T_{decode}(Y)$ is the cost of autoregressively generating an output sequence of length $Y$, and $\Omega$ captures framework overhead from scheduling, serialization, synchronization, and data movement. The inefficiency becomes more severe in multi-agent settings. If $k$ agents consume the same context independently, the system may pay the prefill cost $k$ times:
\begin{equation}
T_{text}^{k} \approx k \cdot T_{prefill}(L) + \sum_{j=1}^{k}T_{decode}(Y_j) + \Omega_{text}.
\end{equation}

\begin{figure}[htpb]
    \begin{center}
    \includegraphics[width=0.6\linewidth]{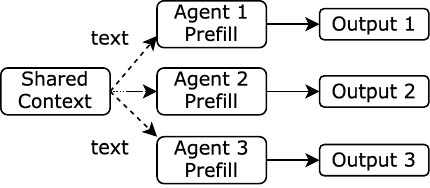}
    \caption{Text-based multi-agent execution duplicates prefill computation across branches that share the same context.} 
    \label{fig:text_duplication}
    \end{center}
\end{figure}

Tree-of-Thought reasoning, multi-agent argument, self-consistency sampling, speculative planning, and parallel retrieval synthesis all exhibit this execution pattern. Although the process in each scenario has a shared prefix or shared context, text-based communication compels each branch to separately recreate the model state. This redundancy is shown in Figure~\ref{fig:text_duplication}. The execution system handles each branch as a separate model invocation even though the agents conceptually share a shared context. Poor branch factor scalability is the outcome, particularly for long-context workloads where time-to-first-token (TTFT) is dominated by prefill.
\subsection{KV Cache as Reusable Execution State}

Serving systems have already demonstrated that performance depends on appropriate KV-cache management. To manage KV cache using block-level paging and minimize memory fragmentation~\cite{kwon2023vllm}, vLLM provides PagedAttention. To enhance KV-cache reuse across structured language-model programs, SGLang introduces RadixAttention~\cite{zheng2023sglang}. These systems show that effective LLM execution depends on the KV cache. Their optimizations, however, mostly function within a serving runtime. A universal distributed abstraction for exposing KV cache as a schedulable state object across multi-agent workflows is not offered by them.

Our work is motivated by this distinction. We don't try to replace the current serving systems. Rather, we investigate the possibility of lifting KV cache from a local serving optimization into a distributed workflow abstraction. According to this perspective, a multi-agent workflow should be able to manage partitions, buffers, and interim outputs in the same manner as distributed data systems may materialize, transmit, fork, reuse, and ultimately evict KV state.

\subsection{Motivating Stateflow Example}

Figure~\ref{fig:text_vs_stateflow} illustrates the difference between textflow and stateflow. Agent B needs to prefill the textual context that Agent A emits in the textflow scenario. Agent A produces a reusable KV state object in addition to a logical output in the stateflow scenario. Then, instead of paying for complete prompt replay, Agent B can resume execution from the transferred state, simply paying for state movement and continuation. By contrasting the cost of state transfer with the cost of text replay, the anticipated benefit may be stated. Text-based execution is beneficially replaced by stateful execution when
\begin{equation}
T_{transfer}(KV) + T_{resume} + \Omega_{state}
<
T_{prefill}(L) + \Omega_{text}.
\label{eq:state_benefit}
\end{equation}
Equation~\ref{eq:state_benefit} defines the central systems condition studied in this paper. Multi-agent workflows can lower TTFT and overall latency by working over state rather than text if transferring and resuming from KV state is less expensive than replaying the prompt.

\section{Stateful Operator Abstraction}

\subsection{Extending the Operator Model}

Each operator is described by AAFLOW~\cite{sarker2026aaflow} as a tuple with inputs, outputs, a transformation function, and a communication pattern. For data-centric workflow steps including embedding, retrieval, reasoning, memory lookup, and index updating, this approach is adequate. It does not, however, specifically differentiate between model execution state and ordinary data. Because KV cache is connected to model settings, token placements, attention layout, device placement, and branch lineage, it is distinct from regular data. Reusing it inappropriately can modify model meaning, while discarding it unnecessarily produces wasteful processing. We therefore extend the original operator definition into a stateful operator:
\begin{equation}
Op_i^{s} =
(I_i, O_i, S_i^{in}, S_i^{out}, f_i, P_i, \sigma_i).
\label{eq:stateful_operator}
\end{equation}
Here, $I_i$ and $O_i$ remain the ordinary data inputs and outputs, while $S_i^{in}$ and $S_i^{out}$ represent input and output state objects. The function $f_i$ describes the logical transformation performed by the operator, and $P_i$ describes the communication pattern used to execute it. The new term $\sigma_i$ denotes the state policy associated with the operator. This policy determines whether state is transferred, aliased, forked, merged, pinned, evicted, or recomputed. Making state movement transparent to the compiler is the goal of Equation~\ref{eq:stateful_operator}. The runtime only observes strings and tensors moving between operators in a text-based workflow. The runtime also detects state dependencies in a stateful workflow, which may affect placement and scheduling. Thus, the execution graph becomes
\begin{equation}
G_s = (V, E_d, E_s),
\end{equation}
where $V$ is the set of operators, $E_d$ is the set of data edges, and $E_s$ is the set of state edges. Data edges describe conventional workflow dependencies, while state edges describe KV-cache reuse and movement.

\subsection{KV State Object}

A KV cache object must carry enough information to determine whether it can be safely reused. We define a state object as
\begin{equation}
S_{KV} = (M, \Theta, B, \Pi, \Lambda, \Gamma),
\label{eq:kv_state_object}
\end{equation}
where $M$ is the model identifier, $\Theta$ is the model and tokenizer configuration, $B$ is the set of KV blocks, $\Pi$ is positional metadata, $\Lambda$ is lineage metadata, and $\Gamma$ describes placement and ownership. The block set $B$ is represented as
\begin{equation}
B = \{b_1,b_2,\dots,b_m\},
\end{equation}
where each block has the form
\begin{equation}
b_j = (K_j,V_j,\ell_j,r_j,d_j).
\end{equation}
The key and value tensors in this representation are $K_j$ and $V_j$, the layer index is $\ell_j$, the token-position range covered by the block is $r_j$, and the device or memory domain where the block is now located is $d_j$. The idea behind this block-level representation is similar to that of paged KV-cache management in vLLM, which manages cache storage in blocks as opposed to a single monolithic sequence~\cite{kwon2023vllm}.

The lineage term $\Lambda$ is particularly important for multi-agent workflows. It records whether a state was produced by direct prefill, by transfer, by fork, or by restricted composition. This prevents the runtime from treating all KV states as interchangeable. Two states are compatible only if they satisfy the predicate
\begin{equation}
Compat(S_a,S_b)=
\mathbb{1}[M_a=M_b]
\cdot
\mathbb{1}[\Theta_a=\Theta_b]
\cdot
\mathbb{1}[\Pi_a \sim \Pi_b],
\end{equation}
where $\Pi_a \sim \Pi_b$ denotes positional compatibility. This predicate is conservative by design: when compatibility cannot be established, the runtime falls back to text replay or recomputation.

\subsection{Stateful Operator Semantics}

The abstraction introduces a family of KV-state operators that extend the original AAFLOW workflow. The first operator is materialization:
\begin{equation}
Op_{kv\_materialize}(x,M) \rightarrow S_{KV}.
\end{equation}
This operator creates a reusable KV state object after consuming a tokenized context $x$ and a model $M$. Materialization distinguishes between the cost of decoding future tokens and the cost of processing the input context, as contrast to regular generation. The second operator is transfer:
\begin{equation}
Op_{kv\_transfer}(S_{KV},n_a,n_b) \rightarrow S'_{KV}.
\end{equation}
A state object is moved or aliased from node $n_a$ to node $n_b$ by this operation. Depending on where the source and destination are located, its communication pattern could be point-to-point transfer, RDMA, or a collective operation. The transfer operator modifies the state's location and ownership metadata but not its logical content.
The third operator is fork:
\begin{equation}
Op_{kv\_fork}(S_{KV},k) \rightarrow \{S_{KV}^{(1)},S_{KV}^{(2)},\dots,S_{KV}^{(k)}\}.
\end{equation}

The abstraction procedure that makes branching effective is called a fork. It generates several logical offspring of a common prefix state. When branches diverge, copy-on-write behavior is applied, even if these descendants may initially share the same physical blocks. Self-consistency sampling, Tree-of-Thought reasoning, multi-agent argument, and parallel plan exploration all directly benefit from this operation. The fourth operator is restricted merge:
\begin{equation}
Op_{kv\_merge}(S_1,S_2,\mu) \rightarrow S^*.
\end{equation}
The merge discipline is specified by the policy $\mu$. Since arbitrary KV-cache merging is not typically semantically correct, we purposefully limit this operation. Merge is only permitted under specific structural limitations, such as prefix-compatible concatenation, segment-aware assembly, or reduction through a summarizing model invocation, because KV state is position-dependent and model-layout-dependent. KV cache is not treated as a general tensor object by this cautious approach.

\subsection{Cost Model for Stateful Execution}

The operator abstraction is useful only if it exposes an optimization target. For text-based execution, the cost of a branch with context length $L$ can be written as
\begin{equation}
T_{text} = T_{prefill}(L) + T_{decode}(Y) + \Omega_{text}.
\end{equation}
For stateful execution, the corresponding cost is
\begin{equation}
T_{state} = T_{transfer}(S_{KV}) + T_{resume} + T_{decode}(Y) + \Omega_{state}.
\end{equation}
The system should choose stateful execution when
\begin{equation}
T_{state} < T_{text}.
\end{equation}
Expanding this condition gives
\begin{equation}
T_{transfer}(S_{KV}) + T_{resume} + \Omega_{state}
<
T_{prefill}(L) + \Omega_{text}.
\label{eq:stateful_condition}
\end{equation}

For a branching workflow with $k$ branches sharing a prefix, the text-based cost is approximately
\begin{equation}
T_{text}^{k}
=
k \cdot T_{prefill}(L)
+
\sum_{j=1}^{k} T_{decode}(Y_j)
+
\Omega_{text}^{k}.
\end{equation}
With KV-state fork, the prefix is materialized once:
\begin{equation}
T_{state}^{k}
=
T_{prefill}(L)
+
T_{fork}(S_{KV},k)
+
\sum_{j=1}^{k} T_{decode}(Y_j)
+
\Omega_{state}^{k}.
\end{equation}
The expected savings therefore come from replacing repeated prefill with a single materialization and a cheaper fork operation:
\begin{equation}
\Delta T
=
(k-1)T_{prefill}(L)
-
T_{fork}(S_{KV},k)
-
(\Omega_{state}^{k}-\Omega_{text}^{k}).
\end{equation}
This equation makes the performance intuition precise. Stateful execution is most beneficial when the shared context is long, the branch factor is high, and state management overhead remains lower than the avoided prefill cost.

\subsection{Correctness and Validity Constraints}

The inability to treat KV cache like arbitrary data is a major problem in stateful abstraction. Only model-specific and sequence-specific requirements make it valid. As a result, three invariants must be maintained in every state transition.

Model compatibility must first be maintained. One incompatible model, tokenizer, or attention layout cannot reuse a KV state created by another. Secondly, positional compatibility needs to be maintained. Fork and merge operations must maintain position semantics since attention state is linked to word order and positional encoding. Third, lineage consistency needs to be maintained. In order for the scheduler to assess if future reuse is safe, a branch formed from a prefix state must document its pedigree. Equation~\ref{eq:kv_state_object} contains metadata terms that encode these requirements. The abstraction does not try unsafe reuse if any invariant fails. Rather, the runtime resorts to text-based recomputation. The abstraction is useful for real systems because of this cautious design, which is crucial for accuracy.


AAFLOW is extended from dataflow to stateflow by the stateful abstraction. The abstraction treats the KV cache as a structured state object with explicit operators, compatibility criteria, placement metadata, and cost models rather than as an opaque local artifact inside the serving engine. This makes it possible for the runtime and compiler to decide when to materialize, transfer, fork, merge, or evict model state. The outcome is a workflow architecture that allows for the effective reuse of costly LLM prefill computation while maintaining the analyzability of operator-driven execution.

\section{Design}
\label{sec:design_and_implementation}
We design a distributed runtime that extends AAFLOW’s operator-driven execution to a \textit{stateful execution system} capable of managing KV cache across multi-agent workflows. The key principle is to treat model execution state as a first-class distributed object, rather than a temporary artifact limited to one node. The runtime integrates a stateful compiler, a distributed KV-state layer, and a state-aware scheduler to coordinate data and state dependencies, as well as transfer/recompute decisions.

\subsection{Stateful Workflow Compilation}

The first step in our system is to extend the compilation process from a purely data-driven model to a hybrid data-state model. Given a workflow $W$, AAFLOW compiles it into an execution graph $G$. We extend this to construct a \textit{stateful execution graph}: $G_s = (V, E_d, E_s)$, where $V$ represents operator nodes, $E_d$ captures data dependencies, and $E_s$ captures state dependencies corresponding to KV cache flow. Unlike traditional DAGs that only encode data movement, \(G_s\) exposes KV-state propagation. State edges allow the scheduler to reason about locality, reuse, and transfer before recomputing shared context. This enables optimizations such as avoiding recomputation when compatible state already exists elsewhere in the system.

\subsection{KV State Representation}

To support distributed state management, we define a structured representation of KV cache. Conceptually, KV cache consists of a sequence of blocks:

\begin{equation}
KV = \{B_1, B_2, \dots, B_n\}
\end{equation}

where each block corresponds to a segment of the input sequence and contains key-value tensors $B_i = (K_i, V_i, l_i, p_i)$. Here, $K_i$ and $V_i$ denote the key and value tensors, $l_i$ identifies the layer index, and $p_i$ encodes the positional range associated with the block. 
We use Apache Arrow~\cite{arrow2016} to represent metadata and block layouts in a columnar, zero-copy format. This allows different components of the system to share state descriptors without serialization overhead. The underlying tensor buffers are transferred using high-performance communication frameworks such as UCX~\cite{ucx2015, staylor2025combining} and MPI~\cite{mpi2005}, ensuring efficient movement across nodes. The transport subsystem overlaps communication and computation, allowing operators to begin execution as KV blocks arrive, improving throughput.

\subsection{KV State Lifecycle}

\begin{figure}[htpb]
    \begin{center}
    \includegraphics[width=0.85\linewidth]{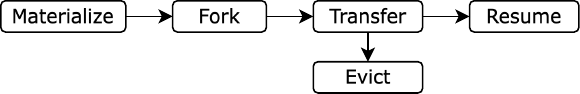}
    \caption{Lifecycle of KV state from creation to reuse and eviction.} 
    \label{fig:kv_lifecycle}
    \end{center}
\end{figure}

The lifecycle of KV state in our system follows a structured sequence, as shown in Figure~\ref{fig:kv_lifecycle}. State is first materialized during the prefill phase of LLM execution, then can be forked to enable parallel reasoning (e.g., in Tree-of-Thought workflows). \textit{Forked} states may be transferred between nodes for distributed processing. When computation resumes, agents use the \textit{transferred}  state directly, avoiding redundant prefill. If necessary, state may be \textit{evicted} due to memory constraints or policy. This lifecycle promotes state reuse, reduces redundant computation, and maintains execution flexibility.

\subsection{Memory Management and Eviction}
Managing KV state for multiple agents strains GPU memory. To mitigate this, a cost-aware eviction strategy balances reuse with memory use. Each state receives a score based on reuse and size. States with lower scores are evicted first under memory pressure, similar to distributed cache strategies but tailored for KV state reuse.

To mitigate GPU memory exhaustion, the runtime implements a heuristic eviction strategy prioritizing KV states based on a combination of memory footprint size and historical reuse frequency. Formalizing and profiling optimal distributed KV eviction policies remains future work.

\subsection{Stateful Workflow Composition}

\begin{figure}[htpb]
    \begin{center}
    \includegraphics[width=0.85\linewidth]{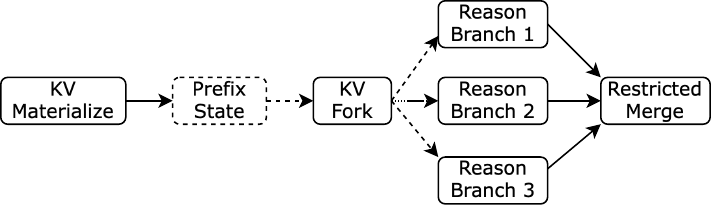}
    \caption{Stateful operator abstraction composes conventional reasoning operators with KV-state operators. Dashed edges represent state dependencies.} 
    \label{fig:stateful_algebra_dag}
    \end{center}
\end{figure}

A stateful workflow composes data operators and state operators into a single graph. A typical branching workflow can be written as

\begin{multline}
Op_{kv\_materialize}
\rightarrow
Op_{kv\_fork}
\rightarrow \\
\{Op_{reason}^{(1)},\dots,Op_{reason}^{(k)}\}
\rightarrow
Op_{kv\_merge}.
\end{multline}
This expression models a workflow where shared context is processed once, branched into multiple reasoning paths, and later merged under specific policies. As illustrated in Figure~\ref{fig:stateful_algebra_dag}, dashed edges indicate KV state flow and solid edges denote standard operator outputs. Explicitly distinguishing state from data dependencies enables the compiler to schedule reasoning branches efficiently, avoiding redundant prefill computation for each branch. Currently, restricted merge supports strictly non-overlapping sequential concatenation of independent branch outputs, avoiding arbitrary tensor-blending of divergent attention states, which would violate positional encoding semantics.

\subsection{State-Aware Scheduling}
The scheduler plays a central role in determining how state is managed during execution. For each operator, the scheduler evaluates whether to transfer existing state or recompute it from scratch. This decision is guided by a cost model:

\begin{equation}
\pi(S) =
\begin{cases}
\text{transfer}, & \text{if } T_{transfer} < T_{prefill} \\
\text{recompute}, & \text{otherwise}
\end{cases}
\end{equation}

Here, $T_{prefill}$ represents the cost of recomputing the state from input text. When transfer is cheaper, the system reuses existing state; otherwise, it falls back to recomputation. More generally, the scheduler optimizes the objective: $\min \sum_i T_i + \lambda \cdot Mem_i$ where $T_i$ is the execution time of operator $i$, $Mem_i$ is its memory footprint, and $\lambda$ controls the trade-off between latency and memory usage. This formulation allows the system to adapt to different workloads and resource constraints. For example, in memory-constrained environments, the scheduler may choose to recompute state rather than store and transfer it.

\subsection{Runtime Integration}

The runtime extends the capabilities of existing LLM serving systems like vLLM~\cite{vllm2023} and SGLang~\cite{sglang2024}, which already manage KV caches efficiently on single nodes, to distributed workflows. Execution involves four key stages: generating the stateful execution graph $G_s$, scheduling operators and state placement, executing operators with KV reuse when possible, and dynamically updating the state graph as states are created or consumed. This approach maintains compatibility with current model-serving infrastructure while supporting advanced distributed optimization.

\section{Implementation}

\subsection{Overview}

\begin{figure}[htpb]
    \begin{center}
    \includegraphics[width=0.85\linewidth]{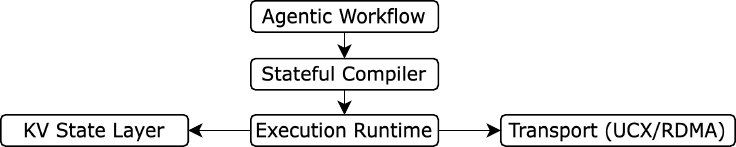}
    \caption{System architecture illustrating the interaction between compiler, runtime, KV state layer, and transport.} 
    \label{fig:stateful_workflow}
    \end{center}
\end{figure}

We implement the proposed stateful agentic abstraction as a distributed runtime that extends AAFLOW with explicit support for KV-state orchestration. The implementation builds on the operator-driven execution model introduced in AAFLOW, where workflows are compiled into communication-aware execution graphs over distributed resources. In contrast to existing agent frameworks, which treat model execution as a black box, our system externalizes KV cache as a manipulable distributed state object. The system consists of four tightly integrated layers: a state-aware compiler, a KV-state manager, a transport subsystem, and an execution runtime (Figure~\ref{fig:stateful_workflow}). These components collectively enable the system to materialize, transfer, reuse, and evict KV state while preserving compatibility with existing LLM serving infrastructures such as vLLM~\cite{kwon2023vllm} and SGLang~\cite{zheng2023sglang}.

\subsection{Build Compiler and Execution Graph}

The compiler extends the workflow of AAFLOW compilation process by incorporating state dependencies into the execution graph. Given a workflow $W$, the compiler constructs a stateful graph $G_s = (V, E_d, E_s)$, where $E_d$ represents data dependencies and $E_s$ represents KV-state dependencies. Each operator is instantiated as:
\begin{equation}
v_i = (Op_i^s, \mathcal{R}_i, \mathcal{L}_i)
\end{equation}

where $\mathcal{R}_i$ captures resource requirements and $\mathcal{L}_i$ encodes locality constraints derived from KV-state placement. During compilation, the system identifies shared context across agents and inserts explicit state operators such as $Op_{kv\_fork}$ and $Op_{kv\_transfer}$. This transformation enables the reuse of model execution state across agents rather than recomputation from text. The compilation process preserves the determinism guarantees of AAFLOW while extending its execution model to include stateflow alongside dataflow.

\subsection{KV-State Manager}

The KV-state manager is responsible for representing and maintaining KV cache across distributed resources. Inspired by block-based KV management in vLLM~\cite{kwon2023vllm}, the system partitions KV cache into fixed-size blocks to enable efficient reuse and transfer. Each KV state is represented as a collection of blocks with associated metadata, as defined in Section 3. Internally, the system maintains a distributed mapping:
\begin{equation}
\mathcal{M}: (state\_id, block\_id) \rightarrow (device, address)
\end{equation}

This mapping allows the runtime to locate KV blocks without scanning global state. The manager also tracks lineage information, ensuring that forked states maintain consistent ancestry relationships. Metadata is encoded using Apache Arrow~\cite{arrow2016}, which provides a columnar, zero-copy representation that can be shared across components without serialization overhead. This design aligns with prior work demonstrating that zero-copy data exchange is critical for high-performance distributed pipelines.

\subsection{Transport Subsystem}

The transport subsystem enables efficient movement of KV state across nodes and devices. It leverages high-performance communication frameworks such as UCX~\cite{ucx2015} and MPI~\cite{mpi2005} to perform zero-copy transfers of tensor buffers. Unlike traditional distributed systems that serialize objects into intermediate formats, our system transfers raw KV blocks directly between memory regions. Metadata is transmitted separately using Arrow descriptors, while large tensor buffers are transferred using RDMA when available. The transfer cost is modeled as:
\begin{equation}
T_{transfer} = \frac{|KV|}{BW} + \delta
\end{equation}

To reduce communication overhead, the system operates at block granularity and transfers only the subset of KV state required by downstream operators. This approach is particularly effective for prefix-based reuse, where only early segments of the sequence are needed. Additionally, the transport subsystem supports overlapping communication with computation. As KV blocks arrive, operators can begin partial execution, improving overall pipeline throughput.

\subsection{Execution Runtime}

The execution runtime orchestrates the execution of the compiled graph across distributed resources. It schedules operators based on both data availability and KV-state availability, ensuring that tasks are executed only when required inputs and state are ready.

Each operator invocation interacts with the underlying LLM serving system through KV-aware APIs. Specifically, when executing a reasoning operator, the runtime injects precomputed KV state into the model, bypassing the prefill stage and directly initiating decoding. This mechanism builds upon the KV reuse capabilities of systems such as vLLM and SGLang, extending them from single-node execution to distributed workflows~\cite{kwon2023vllm,zheng2023sglang}.

\subsection{Fault Tolerance and Consistency}

Maintaining correctness in the presence of distributed state requires careful validation. Each KV state carries metadata that ensures compatibility across operators. In particular, lineage tracking ensures that state is only reused in contexts where it remains semantically valid. When failures occur, the system can recover by recomputing state from the nearest valid prefix. This fallback mechanism ensures robustness without requiring full checkpointing of intermediate states.


The proposed system generalizes agentic execution from a stateless, text-based model to a stateful, distributed model. By explicitly representing, transferring, and reusing KV cache, the system eliminates redundant prefill computation and enables scalable multi-agent workflows. The combination of state-aware compilation, structured state representation, and cost-driven scheduling forms the foundation for efficient stateful execution.

\subsection{Branching Optimization}
Branching workloads are particularly well-suited for stateful execution. Consider a workflow that generates $k$ reasoning branches from a shared prefix. In text-based systems, each branch must independently process the full context:

\begin{equation}
T_{text}^{k} = k \cdot T_{prefill}(L)
\end{equation}

In contrast, our system computes the prefix once and then forks the resulting KV state:

\begin{equation}
T_{state}^{k} = T_{prefill}(L) + k \cdot T_{decode}
\end{equation}

Since decoding is significantly cheaper than prefill, this leads to substantial performance gains. Figure~\ref{fig:stateful_algebra_dag} illustrates this optimization. Restricted merge is used to append non-overlapping segments, rather than attempting to mathematically blend divergent parallel attention matrices, which would corrupt RoPE positional encodings.

\section{Evaluation}
\label{sec:evaluation}

To isolate the performance of distributed KV-transfer from the variance of application-level framework overhead, we evaluate our abstraction using deterministic synthetic prompts and natural questions~\cite{kwiatkowski2019natural} datasets and a trace-driven analytical model parameterized by rigorous empirical microbenchmarks (TTFT, multi-agent computing cost, transfer/recompute tradeoffs, memory footprint, throughput, and framework overhead).
\textbf{AAFLOW+ is also compatible with vLLM, and SGLang backends} 

\subsection{Baselines}

We compare against representative systems from three categories: LLM serving, KV-cache optimization, and multi-agent orchestration.

\textbf{(1) vLLM (PagedAttention)}~\cite{kwon2023vllm}:  
vLLM is a state-of-the-art LLM serving system that optimizes KV cache using block-based memory management and PagedAttention. It represents the strongest baseline for single-node KV reuse and is widely used in production.

\textbf{(2) SGLang (RadixAttention)}~\cite{zheng2023sglang}:  
SGLang improves KV reuse by identifying shared prefixes across structured language programs and exploiting radix-tree-style attention reuse. It provides a natural comparison for prefix-sharing optimizations.

\textbf{(3) DistServe}~\cite{zhong2023distserve}:  
DistServe is a distributed LLM serving system that decouples prefill and decode stages across nodes. While it improves resource utilization, it does not expose KV cache as a reusable distributed object across agents.

\textbf{(4) AAFLOW-text}~\cite{sarker2026aaflow}:  
AAFLOW-text focuses on memory-efficient inference through offloading from RAG and scheduling techniques. It highlights that prefill from vectorstore through RAG could improve, but does not optimize inter-agent KV reuse.

\textbf{(5) Text-based Dense Prefil Orchestration (Sarathi-style)}~\cite{agrawal2023sarathi}:  
This baseline represents current agent frameworks where agents exchange text messages without KV reuse. Each agent performs independent prefill.

\textbf{(6) Communication-oriented KV-cache reuse (KVCOMM)} ~\cite{kvcomm25ye}: 
KVCOMM shares reusable cache context across related agent interactions instead of fully replaying text prompts. Unlike AAFLOW+, it models KV communication without exposing a full workflow-level abstraction for state materialization, fork, transfer, restricted composition, and eviction.


\subsection{Metrics}

We evaluate systems using metrics that capture both model-level performance and system-level efficiency.

\paragraph{Time-to-First-Token (TTFT):} TTFT measures the latency between request arrival and the generation of the first output token:
\begin{equation}
TTFT = T_{prefill} + T_{queue} + \Omega
\end{equation}
Since prefill dominates TTFT for long contexts, reducing redundant prefill is critical.

\paragraph{Aggregate Compute Cost:} The aggregate compute cost is defined as:
\begin{equation}
T_{total} = T_{prefill} + T_{decode} + \Omega
\end{equation}

\paragraph{Framework Overhead ($\Omega$):} Following AAFLOW, we isolate system overhead as:
\begin{equation}
\Omega = T_{total} - (T_{prefill} + T_{decode})
\end{equation}
This includes scheduling, communication, serialization, and synchronization costs.

\paragraph{Throughput:} Throughput is measured as tokens per second:
\begin{equation}
Throughput = \frac{\text{Total tokens generated}}{T_{total}}
\end{equation}

\paragraph{Memory Footprint:} We measure peak KV memory usage:
\begin{equation}
Mem_{KV} = \max_t \sum_i |KV_i(t)|
\end{equation}

\paragraph{KV Reuse Ratio:} To quantify reuse effectiveness, we define:
\begin{equation}
Reuse = \frac{\text{Tokens served from KV cache}}{\text{Total tokens processed}}
\end{equation}

\paragraph{Transfer Efficiency:} We evaluate the ratio between transfer cost and avoided recomputation:
\begin{equation}
Efficiency = \frac{T_{prefill} - T_{transfer}}{T_{prefill}}
\end{equation}

\subsection{Experimental Setup}

All experiments are conducted on a distributed GPU cluster.

\paragraph{Hardware:}
We use a cluster of 4–16 nodes, each equipped with NVIDIA A100 (80GB and 40GB), 32–64 CPU cores, and RDMA-enabled InfiniBand interconnects.

\paragraph{Models and backend:} The HF backend uses \texttt{AutoTokenizer} and
\texttt{AutoModelForCausalLM} with deterministic decoding.  KV metadata is extracted from \texttt{past\_key\_values}; tensors are not serialized by default. The measured real-model values used by the benchmark matrix include prefill time, decode time, TTFT approximation, generated-token count, KV shape metadata,
and KV byte size.

\begin{table}[h]
\centering
\caption{HF, vLLM, SGLang backend use below model coverage in the completed run.}
\label{tab:model_size}
\begin{tabular}{llll}
\toprule
Name & HF model id & Max   & Output  \\
 & HF model id &  context  &  tokens \\
\midrule
Mistral & \texttt{Mistral-7B-Instruct-v0.3} & 32,640 & 64 \\
Llama3 & \texttt{Meta-Llama-3-8B-Instruct} & 8,128 & 64 \\
\bottomrule
\end{tabular}
\end{table}

Mistral was tested up to 32,640 context tokens so that
context\_tokens + output\_tokens stayed within the configured 32,768 token window.  Llama3 was tested up to 8,128 context tokens with 64 output tokens, matching the configured 8,192-token limit (Table - \ref{tab:model_size}). Our runtime integrates with the backend's KV-cache APIs while extending them to support distributed state transfer. Communication is implemented using UCX and NCCL. Mistral-7B, vLLM, and SGLang are licensed under Apache 2.0. Llama-3-8B is licensed under the Meta Llama 3 Community License.

\paragraph{Cost and network parameters:}
For stateful runtime experiments, the configured state-transfer model uses
measured KV bytes.  Experiment 1 used
\texttt{bandwidth\_bytes\_per\_sec = 25,000,000,000}, equivalent to
25 GB/s or 200 Gbps, with \texttt{network\_latency\_sec = 50us},
\texttt{resume\_overhead\_sec = 0.1ms}, and very small framework overheads
(\texttt{omega\_state\_sec = omega\_text\_sec = 50us}).  Experiment 3 sweeps network bandwidth from 10 Gbps to 400 Gbps and uses both RDMA-like 10us and Ethernet-like 100us latency settings.

\subsection{Workloads and Topologies}
The benchmarked scenario is a shared-prefix agentic workflow, not independent
stateless chat.  The main topology is a parallel broadcast / Tree-of-Thought
style DAG:

\begin{multline}
\texttt{root context} \rightarrow \texttt{KV materialize} \rightarrow \texttt{KV fork} \\
\rightarrow \{\texttt{agent}_{1},\ldots,\texttt{agent}_{k}\} \rightarrow \texttt{merge}
\end{multline}

A single root context is processed once, and $k$ downstream agents evaluate different reasoning paths using the same prefix.  The multi-agent scaling experiment sweeps $k \in \{1,2,4,8,16\}$.  The main real-model runs use $Y=64$ generated tokens per branch.  This output length is important: AAFLOW+ primarily reduces prefill and TTFT.  If $Y$ were thousands of tokens, decode would dominate and the relative TTFT advantage would be smaller. The prompts are deterministic synthetic long-context prompts.  They are used to control context length and branch count precisely.  These experiments therefore measure systems behavior---state reuse, transfer, memory, and overhead---rather than dataset answer quality.  

We evaluate three representative workloads: \textit{Multi-Agent Debate:} multiple agents iteratively refine responses using shared context.
\textit{Tree-of-Thought Reasoning:} branching reasoning paths with shared prefixes.
\textit{Retrieval-Augmented Generation (RAG):} agents share retrieved context and perform downstream reasoning. It is not expected to help independent short stateless prompts, because such prompts have no shared state to fork or transfer. Each workload varies in context length (1K–32K tokens), number of agents (2–16), and branching factor (2–16).

\subsection{Meaning of total latency}

The \texttt{total\_latency\_sec} field in these tables is a modeled aggregate workload cost, not the observed Slurm wall-clock runtime of the experiment
script.  For text baselines, the benchmark multiplies measured per-prompt prefill and decode costs by the number of branch instances and prompts.  For example, Experiment 1 uses 16 agents, branch factor 8, and 32 prompts, so dense text passing is charged for $16 \times 8 \times 32 = 4096$ full-context prefill/decode executions.  AAFLOW+ is charged for one shared materialized state plus reused state transfers and continuations.  Therefore large \texttt{total\_latency\_sec} values should be interpreted as aggregate work avoided by state reuse, not as the elapsed time of the Slurm job.

\begin{figure}[htpb]
\centering
\includegraphics[width=\linewidth]{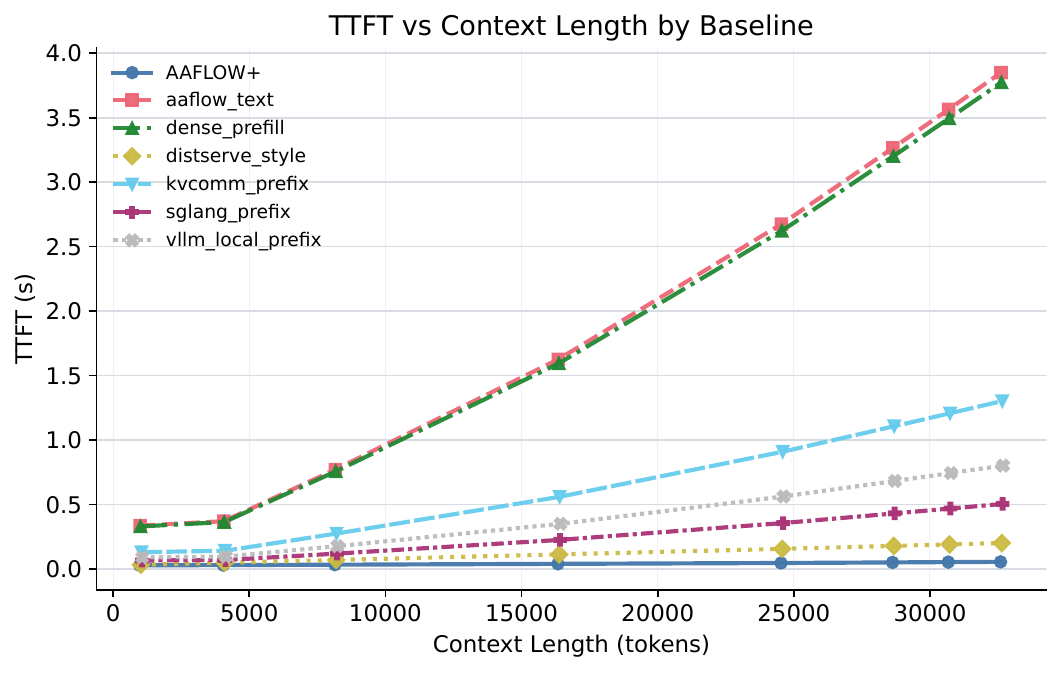}
\captionof{figure}{
Experiment-1: TTFT vs context length across multiple baselines. The curves confirm that text baselines scale with context length, while AAFLOW+ grows slowly because branch cost follows the transfer/resume path rather than repeated prefill
}
\label{fig:ttft-ex1-main}
\end{figure}

\begin{figure}[htpb]
\hfill
\centering
\captionof{table}{Experiment-1: Mean TTFT reduction and Throughput with variation context grid and fixed 16 agents.}
\label{tab:exp1-mean-main}
\resizebox{\linewidth}{!}{
\begin{tabular}{llrr}
\toprule
Model & Baseline & Mean TTFT (s) & Relative to AAFLOW+ \\
\midrule
Mistral & AAFLOW+ & 0.041 & 1.00$\times$ \\
Mistral & dense prefill & 2.017 & 49.2$\times$ slower \\
Mistral & AAFLOW-text & 2.057 & 50.2$\times$ slower \\
Mistral & vLLM local prefix & 0.437 & 10.7$\times$ slower \\
Mistral & SGLang prefix & 0.280 & 6.8$\times$ slower \\
Mistral & KVCOMM & 0.704 & 17.2$\times$ slower \\
Mistral & DistServe style & 0.123 & \textbf{3.0$\times$ slower} \\
\midrule
Llama3 & AAFLOW+ & 0.030 & 1.00$\times$ \\
Llama3 & dense prefill & 0.499 & 16.6$\times$ slower \\
Llama3 & AAFLOW-text & 0.509 & 17.0$\times$ slower \\
Llama3 & vLLM local prefix & 0.124 & 4.1$\times$ slower \\
Llama3 & SGLang prefix & 0.086 & 2.9$\times$ slower \\
Llama3 & KVCOMM & 0.187 & 6.2$\times$ slower \\
Llama3 & DistServe style & 0.052 & \textbf{1.7$\times$ slower} \\
\bottomrule
\end{tabular}
}

\end{figure}

\subsection{Experiment 1: TTFT Reduction}
The first experiment measures time to first token (TTFT) as a function of context length. Text-passing systems repeatedly rebuild the same prefix and pay a TTFT cost that scales with context length: $TTFT_{\mathrm{text}} \propto L.$ AAFLOW+ pays the prefill cost once and then resumes from the distributed state:
\[
TTFT_{\mathrm{state}} \approx T_{\mathrm{transfer}} + T_{\mathrm{resume}}.
\]

Table~\ref{tab:exp1-mean-main}(HF+Mistral, Llama3) and Figure~\ref{fig:ttft-ex1-main}(HF+Mistral)  reports the mean TTFT across all tested context lengths for each model.  AAFLOW+ has the lowest mean TTFT for both models.  On Mistral, dense text prefill, AAFLOW-text and vLLM-local, SGLang, share almost same cold-start TTFT because each must first materialize the prompt prefix locally, while AAFLOW+ has 0.041s, a $49.2\times$ reduction. Local-prefix baselines reduce total repeated work after cache population, which is reflected in aggregate compute cost rather than cold TTFT. The nearest competitor by mean TTFT is DistServe-style at 0.123s, still $2.99\times$ slower than AAFLOW+. The largest contexts show the widening gap most clearly. 

The TTFT curves validate the expected asymptotic behavior.  Dense and text baselines scale with context length because every agent or branch repeats prefix prefill.  AAFLOW+ grows much more slowly because the marginal branch cost is the KV transfer/resume path.  Mistral shows the strongest visible spread because the run reaches 32K context; DistServe-style is the nearest TTFT competitor because it also separates prefill from decode, but it does not expose the same explicit state abstraction and therefore remains slower in these state-transfer workloads.

\begin{figure}[htpb]
\centering
\includegraphics[width=\linewidth]{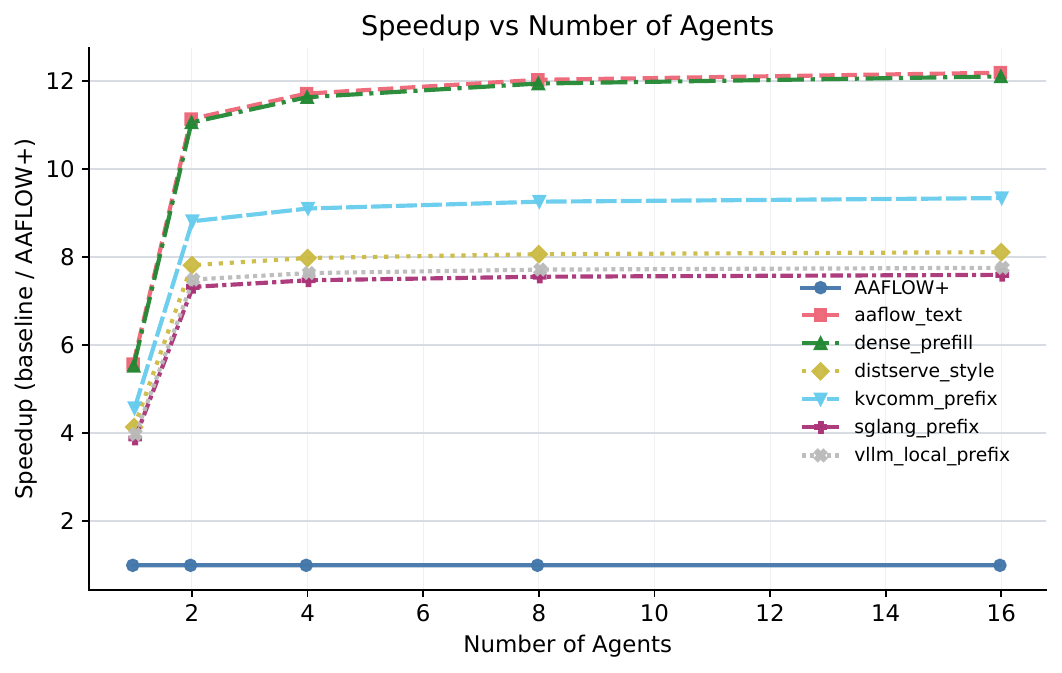}
\captionof{figure}{
Experiment-2: Scaling impact of operator abstraction for multi-agent jobs with HF+Mistral. Total latency and speedup versus the nearest competitor with HF backend. The scaling result is the clearest evidence that the KV state abstraction matters for multi-agent workloads.
}
\label{fig:speedup-main}
\end{figure}

\begin{table}[htpb]
\centering
\captionof{table}{Experiment-2: Aggregate Compute Cost and Efficiency Gain (EG) vs nearest competitor in HF backend}
\label{tab:exp2-speedup-main}
\resizebox{\linewidth}{!}{
\setlength{\tabcolsep}{4pt}
\begin{tabular}{llrrrr}
\toprule
Model & Agents & AAFLOW+ & Dense & SGLang (s) & EG to \\
 &   & (s) & (s)  & Competitor &  SGLang \\
\midrule
Mistral & 1 & 30.700 & 169.757  & 119.135 & 3.88$\times$ \\
Mistral & 2 & 30.700 & 339.514 & 224.850 & 7.32$\times$ \\
Mistral & 4 & 58.371 & 679.027 & 436.281 & 7.47$\times$ \\
Mistral & 8 & 113.714 & 1358.055 & 859.142 & 7.56$\times$ \\
Mistral & 16 & 224.399 & 2716.109 & 1704.865 & 7.60$\times$ \\
\midrule
Llama3 & 1 & 29.067 & 143.959 & 112.396 & 3.87$\times$ \\
Llama3 & 2 & 29.067 & 287.918  & 216.891 & 7.46$\times$ \\
Llama3 & 4 & 56.351 & 575.836 & 425.880 & 7.56$\times$ \\
Llama3 & 8 & 110.918 & 1151.671 & 843.858 & 7.61$\times$ \\
Llama3 & 16 & 220.052 & 2303.342 & 1679.815 & 7.63$\times$ \\
\bottomrule
\end{tabular}
}

\end{table}

\subsection{Experiment 2: Multi-Agent Scaling}
To evaluate multi-agent scaling, we extrapolate aggregate multi-agent compute time using an analytical cost model parameterized by empirical microbenchmarks gathered on our GPU cluster. In text-passing systems, each agent pays for repeated context processing, so cost increases approximately linearly with the agent count. In high-branch-factor workflows (e.g., 16 parallel branches), a single node can't run all decodes without OOM errors, so the workload must be distributed across nodes, requiring KV network transfer. AAFLOW+ materializes the shared context once and then fork the state, which should improve the speedup as $k$ grows:
\[
T^k_{\mathrm{state}} \approx T_{\mathrm{prefill}} + k \cdot T_{\mathrm{decode}},
\]

Table~\ref{tab:exp2-speedup-main} and Figure~\ref{fig:speedup-main} show aggregate compute cost by agent count.  The scaling table shows two effects.  First, AAFLOW+ uses the available stateful parallel width for the first two agents, so 1-agent and 2-agent costs are equal for both models.  Second, after the parallel width is filled, costs grow by waves rather than by full repeated prefill.  Mistral grows from 30.700s at 1--2 agents to 224.399s at 16 agents.  Llama3 grows from 29.067s at 1--2 agents to 220.052s at 16 agents.  The nearest non-AAFLOW+ competitor at every agent count is SGLang prefix, but it remains $3.87\times$--$7.63\times$ slower because local prefix reuse does not remove workflow-level branch duplication.   

The dense baseline is much worse: at 16 agents, the dense total aggregated compute cost is 2716.109 for Mistral. AAFLOW+ converts the agent dimension from repeated prefill into repeated decode continuation from an already materialized state. Even a strong local-prefix baseline cannot fully remove duplicated branch work once many agents need the same prefix. 

\begin{figure}[htpb]
\centering
\includegraphics[width=\linewidth]{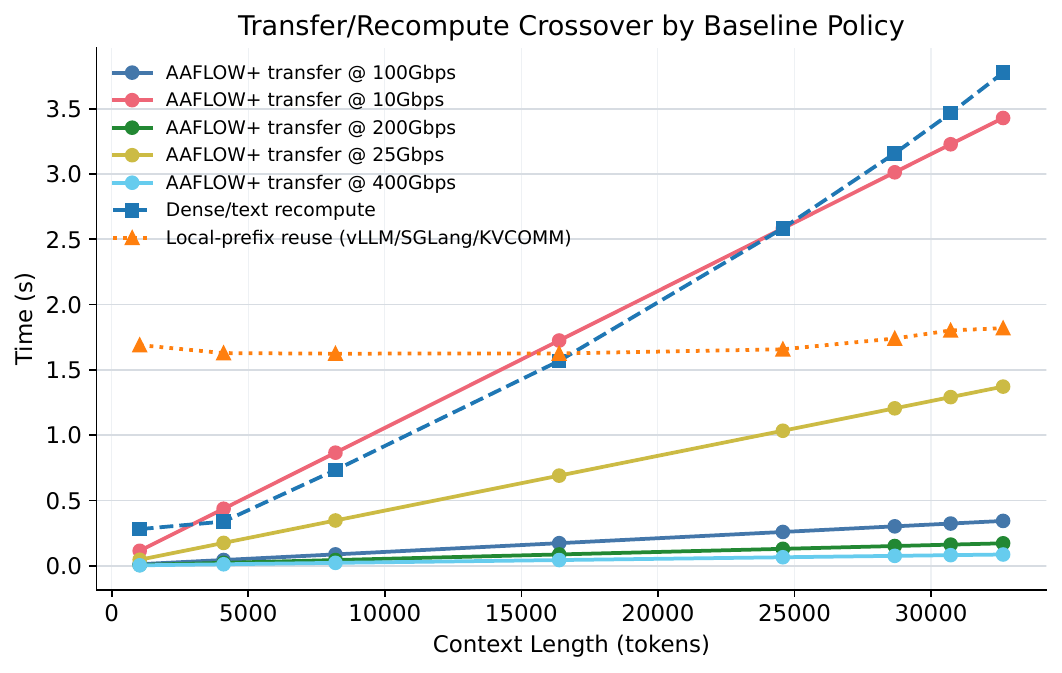}
\captionof{figure}{
Experiment-3: RDMA-like transfer benefit summary with HF backend and Mistral model.  This experiment gives a concrete scheduling rule for stateful execution. The maximum transfer-vs-recompute speedup grows with bandwidth.
}
\label{fig:exp3-transfer_recomputation-main}
\end{figure}

\subsection{Experiment 3: Transfer vs. Recomputation}

The third experiment compares KV transfer cost to recomputing the prompt
prefill where $T_{\mathrm{recompute}} = T_{\mathrm{prefill}}$   :
\[
T_{\mathrm{transfer}} = \frac{\mathrm{KV\ bytes}}{\mathrm{bandwidth}} +
\mathrm{latency},
\]

The scheduler prefer state transfer when $T_{\mathrm{transfer}} < T_{\mathrm{recompute}}$. 
Figure~\ref{fig:exp3-transfer_recomputation-main} summarizes the RDMA-like latency sweep.  
On slow 10 Gbps links, the KV object can be large enough that recomputation may be preferable for some contexts (5 out of 8). This is expected: low bandwidth makes the serialized KV payload expensive.  At 25 Gbps and above, transfer wins at every tested context for both models, measured HF prefill is already more expensive than transferring the measured KV bytes across every tested context. AAFLOW+, therefore, needs both the stateful operators and the cost model: \texttt{KVTransfer} is not always correct, but it is usually correct on high-bandwidth GPU clusters and can be selected explicitly by the scheduler.

\begin{figure}[htpb]

\centering
\includegraphics[width=\linewidth]{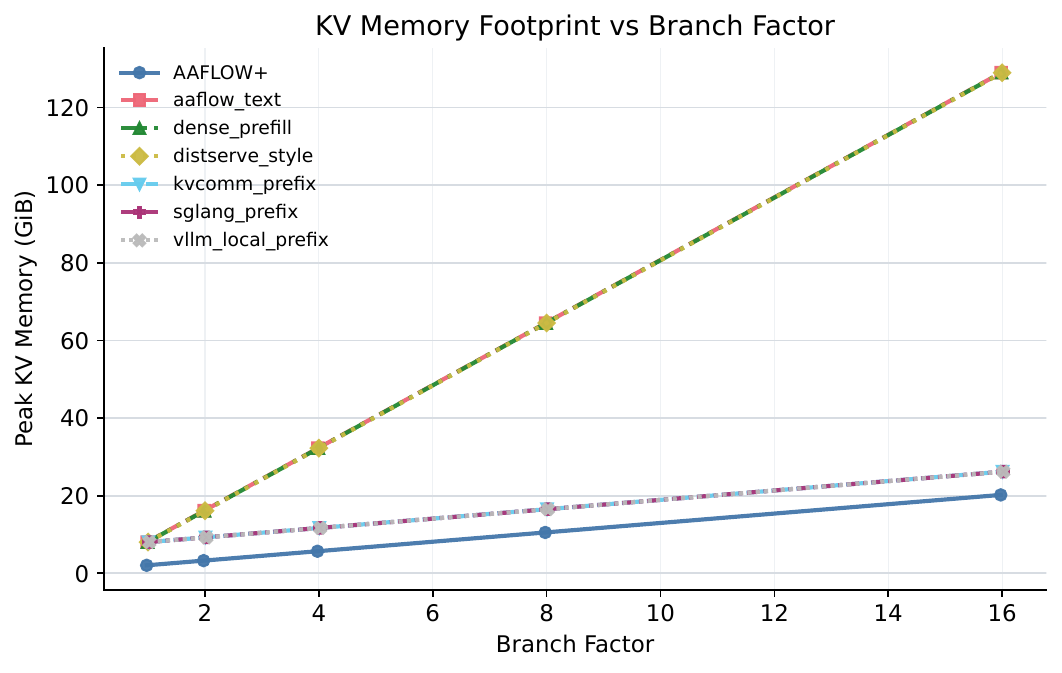}
\captionof{figure}{
Experiment-4: Mean peak KV memory with HF backend and Mistral Model. 
AAFLOW+ reduces memory because it treats KV as a first-class object with lineage and ownership.
}
\label{fig:exp4-mean-main}

\end{figure}

\begin{table}

\centering
\captionof{table}{Experiment-4: Mean peak KV memory and memory ratio with HF backend}
\label{tab:exp4-mean-main}
\resizebox{\linewidth}{!}{
\begin{tabular}{llrr}
\toprule
Model & Baseline & Peak KV& Memory \\
 &  & (GiB) & Ratio \\
\midrule
Mistral & AAFLOW+ & 8.355 & 1.00$\times$ \\
Mistral & dense prefill & 49.987 & 5.98$\times$ larger \\
Mistral & AAFLOW-text & 53.986 & 6.46$\times$ larger \\
Mistral & vLLM local prefix & 14.351 & \textbf{1.72$\times$ larger} \\
Mistral & SGLang prefix & 14.591 & 1.74$\times$ larger \\
Mistral & KVCOMM & 15.702 & 1.88$\times$ larger \\
Mistral & DistServe style & 50.995 & 6.10$\times$ larger \\
\midrule
Llama3 & AAFLOW+ & 4.210 & 1.00$\times$ \\
Llama3 & dense prefill & 25.188 & 5.98$\times$ larger \\
Llama3 & AAFLOW-text & 27.202 & 6.46$\times$ larger \\
Llama3 & vLLM local prefix & 7.231 & \textbf{1.72$\times$ larger} \\
Llama3 & SGLang prefix & 7.311 & 1.73$\times$ larger \\
Llama3 & KVCOMM & 7.912 & 1.88$\times$ larger \\
Llama3 & DistServe style & 25.695 & 6.10$\times$ larger \\

\bottomrule
\end{tabular}
}

\end{table}

\subsection{Experiment 4: Memory Efficiency}
The fourth experiment compares peak KV memory footprint across baselines.  The stateful design should avoid allocating redundant prefix KV for every branch. AAFLOW+ stores one materialized prefix plus branch continuation state; dense and text baselines effectively duplicate the prefix across branches. Table~\ref{tab:exp4-mean-main} and Figure~\ref{fig:exp4-mean-main} reports mean peak KV memory across branch factors 1, 2, 4, 8, and 16.  For Mistral, AAFLOW+ averages 8.355 GiB peak KV memory. Dense, AAFLOW-text, and DistServe-style average 49.987 GiB, or $5.98\times$ the AAFLOW+ memory.  vLLM-local and SGLang average 14.351 GiB ($1.72\times$ AAFLOW+), and 14.591 GiB ($1.74\times$ AAFLOW+), while KVCOMM averages 15.702 GiB ($1.88\times$). The nearest competitor varies with the branch factor.  At branch factors 1, 2, and 4, KVCOMM is closest to AAFLOW+. At branch factors 8 and 16, vLLM-local becomes closest.  Even then, AAFLOW+ uses less memory at every branch factor.

Forked states initially share block references, so memory grows with branch continuation rather than with full-context duplication. The vLLM-local and SGLang profiles also reduce duplicate prefix storage, but they are local-prefix baselines rather than distributed state-object schedulers. SKVCOMM captures communication-oriented prefix reuse but has a lower reuse fraction in this benchmark. The result is that AAFLOW+ has the lowest peak memory in every tested branch-factor setting.

\subsection{Experiment 5: Throughput and Framework Overhead}

The fifth experiment measures effective throughput and framework overhead
$\Omega$ shown in Table~\ref{tab:exp5-main}.  AAFLOW+ should improve throughput by removing repeated prefill work and reducing text serialization.  In the benchmark schema, throughput is reported as generated tokens per second, and $\Omega$ captures framework-side scheduling, serialization, and orchestration overhead modeled by the baseline adapters. AAFLOW+ achieves approximately 302 tokens/s on Mistral.  The nearest throughput competitor is vLLM-local/SGLang/DistServe-style at about 38--39 tokens/s.  The resulting throughput advantage is about $7.63\times$--$8.04\times$ against those nearest competitors and roughly $9.47\times$ against dense text prefill, although dense text prefill has low overhead.

\begin{table}[htpb]
\centering
\captionof{table}{Experiment-5: Mean throughput and framework overhead with fixed context grid and variation agents.}
\label{tab:exp5-main}
\resizebox{\linewidth}{!}{
\begin{tabular}{llrr}
\toprule
Model & Baseline & Throughput & $\Omega$ (s) \\
 &  & (tok/s) &\\
\midrule
Mistral & AAFLOW+ & \textbf{302.61} & 0.0075 \\
Mistral & dense prefill & 31.94 & \textbf{0.0019} \\
Mistral & AAFLOW-text & 31.84 & 11.1659 \\
Mistral & vLLM local prefix & 38.81 & 0.0627 \\
Mistral & SGLang prefix & 39.62 & 0.0657 \\
Mistral & KVCOMM & 35.59 & 21.6770 \\
Mistral & DistServe style & 37.62 & 0.0597 \\
\midrule
Llama3 & AAFLOW+ & \textbf{300.57} & 0.0075 \\
Llama3 & dense prefill & 33.61 & \textbf{0.0019} \\
Llama3 & AAFLOW-text & 33.54 & 6.7272 \\
Llama3 & vLLM local prefix & 38.46 & 0.0627 \\
Llama3 & SGLang prefix & 39.27 & 0.0657 \\
Llama3 & KVCOMM & 36.16 & 13.0375 \\
Llama3 & DistServe style & 37.46 & 0.0597 \\

\bottomrule
\end{tabular}
}

\end{table}

The throughput numbers are consistent with the aggregated compute costs experiments.  AAFLOW+ executes one measured prefill and then reuses state; text baselines repeatedly pay the full context construction and prefill cost.  The explicit abstraction also keeps framework overhead small in this simulation profile.  KVCOMM has a meaningful modeled communication overhead in these runs, which explains the larger $\Omega$ values and lower throughput relative to AAFLOW+.

\section{Related Work}

\textbf{Distributed Data Systems and Operator Abstractions:} The design of AAFLOW~\cite{sarker2026aaflow} and our extension builds upon a long line of work in distributed data systems, where computation is expressed as compositions of operators over structured data. Systems such as MapReduce~\cite{dean2008mapreduce}, Spark~\cite{zaharia2016spark}, and Flink~\cite{carbone2015flink} demonstrate how declarative data transformations can be compiled into efficient distributed execution plans. More recent frameworks such as Modin~\cite{petersohn2020modin} and Cylon~\cite{abeykoon2020cylon} extend this paradigm to high-performance dataframes, emphasizing communication-aware execution and parallel processing patterns~\cite{perera2023supercharging, perera2023depth, shan2022hybrid}. AAFLOW adapts these ideas to agentic workflows by treating embedding, retrieval, reasoning, and memory operations as composable operators. Our work further extends this abstraction by introducing stateful operators over KV cache, effectively generalizing dataflow systems into stateflow systems.

\textbf{Workflow Runtimes and Distributed Execution:} A large body of work has explored distributed workflow execution and task scheduling. Systems such as Ray~\cite{moritz2018ray}, Dask~\cite{rocklin2015dask}, and Parsl~\cite{babuji2019parsl} provide flexible task-based execution frameworks for parallel and distributed workloads. HPC-oriented systems such as RADICAL-Pilot~\cite{merzky2022radical, sarker2024radical, sarker2024design, osti_10634837} and Pegasus~\cite{deelman2015pegasus} focus on large-scale scientific workflows. While these systems provide powerful execution substrates, they treat tasks as independent units and do not explicitly model LLM-specific state such as KV cache. As a result, they cannot exploit opportunities for state reuse across agents. Our work complements these systems by introducing an abstraction that exposes state dependencies and enables scheduling decisions based on both data and model state.

\textbf{LLM Programming and Agent Frameworks:} Recent work on LLM programming has focused on improving the expressiveness and composability of language model applications. Frameworks such as LangChain~\cite{langchain2023}, LangGraph~\cite{langgraph2024}, and AutoGen~\cite{wu2023autogen} enable multi-agent workflows, tool invocation, and iterative reasoning. DSPy~\cite{khattab2023dspy} further proposes a declarative programming model for optimizing LLM pipelines. These systems treat orchestration as a high-level programming problem but largely ignore the underlying systems cost of data movement and state recomputation. In particular, they rely on text-based communication between agents, leading to repeated prefill computation. Our work differs by introducing a systems-level abstraction that allows agents to exchange execution state directly.

\textbf{LLM Serving and KV-Cache Optimization:} A growing body of work has focused on optimizing LLM inference through efficient KV-cache management. vLLM~\cite{kwon2023vllm} introduces PagedAttention, which uses block-based memory allocation to improve throughput and reduce fragmentation. SGLang~\cite{zheng2023sglang} proposes RadixAttention, enabling prefix sharing across structured programs. FlexGen~\cite{sheng2023flexgen} explores memory-compute trade-offs through offloading strategies, while DistServe~\cite{zhong2023distserve} and LMCache~\cite{liu2025lmcache} separate prefill and decode stages to improve resource utilization. These systems demonstrate that KV cache is a critical performance factor in LLM serving. However, their optimizations are confined to single-node or single-request execution contexts with efficient storage, movement, and management of KV cache across heterogeneous memory tiers and serving backends. They do not expose KV cache as a distributed systems object that can be shared across multiple agents. Our work extends KV-cache optimization from local serving to distributed multi-agent workflows rather than focusing solely on cache management.

\textbf{Distributed and Multi-Tenant LLM Serving:} Recent work has explored scaling LLM serving across multiple tenants and distributed environments. Systems such as Orca~\cite{yu2022orca} and Sarathi~\cite{agrawal2023sarathi} improve GPU utilization through scheduling and batching techniques. Helix~\cite{li2023helix} focuses on multi-tenant inference with efficient resource sharing. While these systems address throughput and fairness, they do not explicitly consider KV-state reuse across independent requests or agents. Our approach is orthogonal, focusing on reducing redundant computation by reusing state rather than improving scheduling alone.

\textbf{Retrieval-Augmented Generation and Memory Systems:} Retrieval-Augmented Generation (RAG) systems integrate external knowledge with LLM inference to improve factual accuracy~\cite{lewis2020rag}. Recent work has extended this paradigm with more sophisticated memory and retrieval mechanisms, such as MemoRAG~\cite{wang2024memorag}, HippoRAG~\cite{yang2024hipporag}, and CueRAG~\cite{gao2024cuerag}. These systems demonstrate the importance of persistent memory for multi-step reasoning. However, most RAG systems treat memory as external data rather than internal model state. They optimize retrieval and indexing but do not address the cost of reprocessing retrieved context within the model. Our work complements these approaches by enabling the reuse of the internal KV representation of retrieved context.

Existing research has focused on two areas: improving LLM efficiency via KV-cache optimization and enhancing workflow expressiveness with agent frameworks. We unify these by treating the KV cache as a distributed state within an operator abstraction, enabling optimizations like cross-agent state reuse, distributed KV transfer, and state-aware scheduling. Extending AAFLOW’s operator-driven model with stateflow lays the groundwork for scalable, efficient, and reproducible multi-agent LLM systems.
\section{Limitations and Future Work}
These results should be viewed within defined limits. 
The workloads use synthetic, deterministic prompts \cite{kwiatkowski2019natural}, so answer quality is not evaluated against external datasets. The main output length is $Y=64$ tokens; longer outputs would reduce prefill reuse benefits and place greater emphasis on decoding. AAFLOW+ targets shared-prefix multi-agent workflows, not independent stateless inference. Infrastructure assessments are based on controlled experiments rather than production cluster data. We define architectural objectives for eviction and memory-constrained scheduling, but currently evaluate only unconstrained memory. 

AAFLOW+ opens several directions for future research. First, extending the workflow to support \textbf{heterogeneous model environments} remains an important challenge. Current compatibility constraints assume identical model architectures and positional encodings, but real-world systems may involve mixtures of models, adapters, or fine-tuned variants. Generalizing state reuse across heterogeneous settings would broaden applicability. Second, \textbf{cost-aware scheduling} can be further developed. While this work demonstrates the tradeoff between KV transfer and recomputation, future systems could incorporate dynamic bandwidth estimation, workload prediction, and adaptive placement strategies to optimize execution decisions in real time. Third, integrating \textbf{stateful execution with emerging model architectures} is a promising direction. While this work focuses on transformer-based KV cache, alternative architectures such as state space models introduce different notions of internal state. Extending the proposed abstraction to support these models could enable a unified treatment of execution state beyond KV representations. Fourth, \textbf{fault tolerance and persistence} for stateful workflows require deeper investigation. Although recomputation provides a fallback mechanism, efficient checkpointing, partial state recovery, and durable storage of KV state remain open problems for large-scale deployments. Finally, future work can explore \textbf{programming abstractions and developer tooling} for state-aware workflows. Providing higher-level interfaces, debugging tools, and visualization systems for stateflow execution would make these capabilities accessible to a broader class of applications.


\section{Conclusion}

We presented \textbf{AAFLOW+}, a stateful extension of agentic workflow abstraction that makes KV cache a first-class distributed systems object in AI memory instead of just a local inference optimization. AAFLOW+ presents a \emph{stateflow} abstraction that allows agents to directly interchange and reuse execution state, in contrast to current multi-agent LLM systems that rely on text-based communication and frequently recalculate shared context. Our method exposes state reuse at the workflow level by defining operators for KV-state materialization, transfer, fork, restricted composition, and eviction, and assembling workflows into communication-aware graphs. The runtime externalizes KV cache using explicit metadata descriptors and zero-copy communication pathways, enabling efficient state reuse across distributed agents while preserving correctness through compatibility constraints. Experimental results on Mistral-7B and Llama-3-8B show that AAFLOW+ significantly improves efficiency, achieving up to \textbf{50.2$\times$} reduction in TTFT, \textbf{7.63$\times$} lower multi-agent latency at 16-agent scale, \textbf{1.72$\times$--6.10$\times$} reduction in peak KV memory, and over \textbf{7.74$\times$} improvement in throughput. These findings show that duplicate computation and framework overhead in multi-agent LLM execution can be significantly reduced by substituting explicit KV-state sharing for text forwarding in AI memory.

\bibliographystyle{ACM-Reference-Format}
\bibliography{aaflowplus}
\newpage
\appendix
\section*{Appendix}
\addcontentsline{toc}{section}{Appendix}
\appendix

\section{Extended System Design}
\label{app:extended_design}

This appendix expands the design and implementation details of AAFLOW+. The main paper presents the operator abstraction, state object, scheduler, and evaluation results. Here, we describe how the compiler, KV-state manager, transport layer, runtime, and backend adapters interact during execution.

\subsection{Design Components}

AAFLOW+ is organized around five cooperating components: the stateful compiler, the KV-state manager, the transport subsystem, the state-aware scheduler, and the execution runtime. The stateful compiler converts an agentic workflow into a graph containing both data edges and state edges. The KV-state manager tracks block-level KV metadata, ownership, lineage, and placement. The transport subsystem moves KV blocks across devices and nodes using Arrow metadata and UCX/RDMA-style communication. The scheduler decides whether to transfer, reuse, fork, evict, or recompute state. Finally, the execution runtime coordinates backend invocation through Hugging Face, vLLM, or SGLang-compatible execution paths.

Given a workflow $W$, the compiler constructs a stateful execution graph
\[
G_s=(V,E_d,E_s),
\]
where $V$ is the set of operator instances, $E_d$ represents ordinary data dependencies, and $E_s$ represents KV-state dependencies. Each vertex is annotated as
\[
v_i = (Op_i^s, R_i, L_i, C_i),
\]
where $Op_i^s$ is the stateful operator, $R_i$ describes resource requirements, $L_i$ captures locality constraints derived from KV-state placement, and $C_i$ stores compatibility constraints such as model identity, tokenizer configuration, attention layout, and positional encoding.

The compiler inserts KV-state operators when it detects shared prompt prefixes or branchable contexts. For example, a text-centric workflow
\[
Context \rightarrow \{Agent_1,\ldots,Agent_k\}
\]
is rewritten as

\begin{multline}
Op_{kv\_materialize}(Context)
\rightarrow Op_{kv\_fork}(S_{KV},k) \\
\rightarrow  \{Agent_1(S_{KV}^{(1)}),\ldots,Agent_k(S_{KV}^{(k)})\}.
\end{multline}

This transformation makes prefix reuse explicit and allows the runtime to schedule reasoning branches without forcing each branch to replay the same context.

\subsection{KV-State Metadata Catalog}

AAFLOW+ maintains a metadata catalog for every materialized KV state. The catalog separates logical state identity from physical tensor placement. This allows the scheduler to reason about state reuse without scanning device memory or reconstructing prompt text. Each KV state is represented as
\[
S_{KV}=(M,\Theta,B,\Pi,\Lambda,\Gamma),
\]
where $M$ is the model identifier, $\Theta$ is the model and tokenizer configuration, $B$ is the set of KV blocks, $\Pi$ stores positional metadata, $\Lambda$ records lineage, and $\Gamma$ records placement and ownership. Internally, the KV-state manager maintains the mapping
\[
\mathcal{M}:(state\_id, block\_id)\rightarrow(device,node,address,range,owner).
\]

\begin{table}[htpb]
\centering
\caption{Metadata maintained by the AAFLOW+ KV-state manager.}
\label{tab:kv_metadata_catalog}
\begin{tabularx}{0.48 \textwidth}{| l | X |}
\toprule
Field & Purpose \\
\midrule
\texttt{state\_id} & Unique logical identifier for a KV state. \\
\texttt{model\_id} & Model family and checkpoint used to create the state. \\
\texttt{tokenizer\_id} & Tokenizer and vocabulary compatibility. \\
\texttt{position\_range} & Token positions by each KV block. \\
\texttt{layer\_id} & Transformer layer with the KV block. \\
\texttt{block\_id} & Block identifier for partial transfer and reuse. \\
\texttt{lineage} & Materialize, fork, transfer, merge, or evict history. \\
\texttt{device} & GPU, CPU, or remote memory location. \\
\texttt{owner} & Runtime worker currently responsible for the block. \\
\texttt{refcount} & Number of logical states sharing the block. \\
\texttt{last\_access} & Timestamp or logical clock for eviction. \\
\bottomrule
\end{tabularx}
\end{table}

This design allows multiple forked states to share the same physical prefix blocks. When a branch diverges, the runtime applies copy-on-write behavior: shared prefix blocks remain aliased, while branch-specific continuation blocks are allocated separately.

\section{Use Case: Collaborative RAG with Shared Evidence}
\label{app:use_case_rag}

A representative use case is collaborative retrieval-augmented generation over a long document collection. Suppose a user asks a complex scientific question requiring retrieval, risk analysis, verification, and final synthesis. A conventional multi-agent pipeline may use four agents: a retriever, a reasoning agent, a verifier, and a summarizer. All agents require the same retrieved evidence prefix, but they perform different downstream tasks.

In a text-centric implementation, the retriever emits passages as text. Each downstream agent then receives the same passages and performs a separate prefill over the identical context:
\[
T_{\mathrm{text}}^k \approx k\cdot T_{\mathrm{prefill}}(L)+
\sum_{j=1}^{k}T_{\mathrm{decode}}(Y_j)+\Omega_{\mathrm{text}}.
\]
This duplicates both compute and KV memory.

With AAFLOW+, the retrieved evidence is materialized once as KV state:
\[
S_{KV}=Op_{kv\_materialize}(Context,M).
\]
The runtime then forks this state:
\[
\{S_{KV}^{(1)},\ldots,S_{KV}^{(k)}\}
=
Op_{kv\_fork}(S_{KV},k),
\]
and assigns the forked states to the reasoning, verification, and synthesis agents. Each agent resumes from the shared context state and only pays the cost of branch-specific continuation:
\[
T_{\mathrm{state}}^k
\approx T_{\mathrm{prefill}}(L)
+
T_{\mathrm{fork}}(S_{KV},k)
+
\sum_{j=1}^{k}T_{\mathrm{decode}}(Y_j)
+
\Omega_{\mathrm{state}}.
\]
The benefit is largest when the retrieved context is long, the number of agents is large, and the branch outputs are short relative to the shared prefix.

\section{End-to-End Execution}
\label{app:end_to_end_call_flow}

This section describes the full AAFLOW+ call flow for a shared-prefix multi-agent workflow.

\textbf{Workflow Submission:} The user submits a workflow containing a shared context and multiple downstream agents:
\[
W=(Context,\{Agent_1,\ldots,Agent_k\},Policy).
\]
The policy specifies whether state reuse is allowed, whether branches can share physical blocks, and whether transfer or recomputation should be preferred under memory pressure.

\textbf{Compilation:} The compiler parses the workflow and constructs the stateful graph:
\[
G_s=(V,E_d,E_s).
\]
It identifies common prefixes, inserts $Op_{kv\_materialize}$ before the first shared reasoning stage, inserts $Op_{kv\_fork}$ before branch expansion, and annotates each branch with state compatibility constraints.

\textbf{State Materialization:} The runtime sends the shared context to the backend model and executes the prefill stage once. The backend produces KV tensors, which are registered with the KV-state manager:
\[
Context \xrightarrow[]{prefill} S_{KV}.
\]
The KV-state manager records block identifiers, token ranges, layer identifiers, placement, and lineage.

\textbf{Scheduling and Placement:} The scheduler chooses placement for each branch. For a branch assigned to the same node as the materialized KV state, the scheduler prefers aliasing or local reuse. For a branch assigned to a remote node, it compares transfer cost against recomputation:
\[
\pi(S)=
\begin{cases}
transfer, & T_{\mathrm{transfer}}(S)<T_{\mathrm{prefill}}(L),\\
recompute, & otherwise.
\end{cases}
\]

\textbf{Execution:} Agents resume from the assigned KV state and generate branch-specific continuations. The runtime updates the state graph as new continuation blocks are created:
\[
S_{KV}^{(j)} \rightarrow Decode_j(Y_j) \rightarrow Output_j.
\]

\textbf{Merge and Finalization:} The outputs are merged using a safe reduction policy. AAFLOW+ does not arbitrarily blend divergent attention states. Instead, it supports restricted merge patterns such as non-overlapping sequential concatenation, prefix-compatible extension, or text-level reduction through a summarizing model invocation.

\section{Algorithms for KV Operators}
\label{app:kv_operation_call_flows}

Each KV state follows a finite-state lifecycle:
\begin{multline}
Created \rightarrow Registered \rightarrow Active \rightarrow Forked \\
\rightarrow Transferred \rightarrow Resumed \rightarrow Evicted.
\end{multline}

\begin{table*}[htpb]
\centering
\caption{AAFLOW+ KV-state lifecycle transitions.}
\label{tab:kv_state_transitions}
\begin{tabular}{lll}
\toprule
Transition & Trigger & Runtime Action \\
\midrule
Created $\rightarrow$ Registered & Backend prefill completes & Insert metadata into catalog. \\
Registered $\rightarrow$ Active & Scheduler assigns owner & Pin or map blocks to worker. \\
Active $\rightarrow$ Forked & Branch expansion & Create child state descriptors. \\
Forked $\rightarrow$ Transferred & Remote branch placement & Transfer required blocks. \\
Transferred $\rightarrow$ Resumed & Agent execution begins & Inject state into backend. \\
Resumed $\rightarrow$ Active & Decode creates continuation & Register new branch blocks. \\
Active $\rightarrow$ Evicted & Memory pressure & Free low-priority blocks. \\
\bottomrule
\end{tabular}
\end{table*}

The lifecycle is maintained by the KV-state manager and updated after every operator invocation, shown in Table-\ref{tab:kv_state_transitions}. Because lineage is explicit, the scheduler can distinguish original materialized prefixes, forked aliases, transferred blocks, and branch-specific continuation state.

\subsection{KV Materialization}

KV materialization converts a text context into reusable execution state. This operation is invoked once for a shared prefix.

\begin{algorithm}[htpb]
\caption{\textsc{KV-Materialize}}
\label{alg:kv_materialize}
\begin{algorithmic}[1]
\Require Context tokens $x$, model $M$, tokenizer/configuration $\Theta$
\Ensure KV state $S_{KV}$
\State Validate model and tokenizer configuration.
\State Execute backend prefill on $x$ using $M$.
\State Collect layer-wise KV tensors $\{(K_{\ell},V_{\ell})\}$.
\State Partition tensors into KV blocks $B=\{b_1,\ldots,b_m\}$.
\State Create metadata $(M,\Theta,\Pi,\Lambda,\Gamma)$.
\State Register $(state\_id,block\_id)\rightarrow(device,node,address)$ in the KV-state manager.
\State Return $S_{KV}=(M,\Theta,B,\Pi,\Lambda,\Gamma)$.
\end{algorithmic}
\end{algorithm}

Materialization is the only operation that pays the full prefill cost for the shared prefix. All downstream reuse is expressed as fork, transfer, alias, or resume.

\subsection{KV Fork}

KV fork creates multiple logical descendants from a shared prefix state.

\begin{algorithm}[htpb]
\caption{\textsc{KV-Fork}}
\label{alg:kv_fork}
\begin{algorithmic}[1]
\Require Parent state $S_{KV}$, number of branches $k$
\Ensure Forked states $\{S_{KV}^{(1)},\ldots,S_{KV}^{(k)}\}$
\State Check that $S_{KV}$ is valid and not evicted.
\For{$j=1$ to $k$}
    \State Create child metadata with lineage $\Lambda^{(j)}=\Lambda \cup \{fork(j)\}$.
    \State Alias parent prefix blocks by increasing block reference counts.
    \State Assign branch-specific ownership and placement policy.
\EndFor
\State Return $k$ logical child states.
\end{algorithmic}
\end{algorithm}

Fork does not copy the entire KV tensor by default. It creates logical children that share prefix blocks until branch-specific continuation requires new allocation.

\subsection{KV Transfer}

KV transfer moves or aliases KV blocks between workers. The scheduler invokes transfer only when it is cheaper than recomputation.

\begin{algorithm}[htpb]
\caption{\textsc{KV-Transfer}}
\label{alg:kv_transfer}
\begin{algorithmic}[1]
\Require State $S_{KV}$, source node $n_a$, destination node $n_b$
\Ensure Destination state $S'_{KV}$
\State Estimate transfer cost $T_{\mathrm{transfer}}=|KV|/BW+\delta$.
\State Estimate recomputation cost $T_{\mathrm{prefill}}(L)$.
\If{$T_{\mathrm{transfer}}\geq T_{\mathrm{prefill}}(L)$}
    \State Return \textsc{RecomputeDecision}.
\EndIf
\State Select required KV blocks based on downstream position range.
\State Send Arrow metadata descriptor to $n_b$.
\State Transfer KV tensor buffers using UCX/RDMA/MPI transport.
\State Register received blocks in the destination KV-state manager.
\State Update placement metadata $\Gamma$ and lineage $\Lambda$.
\State Return transferred state $S'_{KV}$.
\end{algorithmic}
\end{algorithm}

The operation works at block granularity, allowing the runtime to transfer only the prefix segments needed by the downstream branch.

\subsection{KV Resume}

KV resume injects precomputed state into a backend invocation and skips repeated prefill.

\begin{algorithm}[htpb]
\caption{\textsc{KV-Resume}}
\label{alg:kv_resume}
\begin{algorithmic}[1]
\Require State $S_{KV}$, branch prompt suffix $q$, model backend $B$
\Ensure Generated output $y$
\State Validate compatibility of $S_{KV}$ with backend model and tokenizer.
\State Load or map KV blocks into the backend-visible memory domain.
\State Initialize decoding position from $\Pi$.
\State Append branch-specific suffix tokens $q$ if required.
\State Decode output tokens using reused KV state.
\State Register any newly generated continuation KV blocks.
\State Return generated output $y$.
\end{algorithmic}
\end{algorithm}

Resume is the operation that converts state reuse into TTFT reduction. Instead of rebuilding the shared prefix, the backend continues from the imported or mapped KV state.

\subsection{Restricted KV Merge}

Restricted merge reconciles branch outputs without unsafe tensor-level blending of divergent KV states.

\begin{algorithm}[htpb]
\caption{\textsc{Restricted-KV-Merge}}
\label{alg:kv_merge}
\begin{algorithmic}[1]
\Require States or outputs $\{S_1,\ldots,S_k\}$, merge policy $\mu$
\Ensure Merged state or merged text output
\State Check model compatibility and positional compatibility.
\If{$\mu$ permits prefix-compatible concatenation}
    \State Concatenate non-overlapping compatible state segments.
\ElsIf{$\mu$ requires semantic reduction}
    \State Convert branch outputs to text and call a summarizing/reduction operator.
\Else
    \State Reject tensor-level merge and fall back to text-level reduction.
\EndIf
\State Record merge lineage in $\Lambda$.
\State Return merged result.
\end{algorithmic}
\end{algorithm}

This conservative merge rule is necessary because KV state is position-dependent and model-layout-dependent. AAFLOW+ therefore treats merge as a controlled operator rather than as arbitrary tensor composition.

\subsection{KV Eviction}

KV eviction releases state when memory pressure exceeds the configured threshold.

\begin{algorithm}[htpb]
\caption{\textsc{KV-Evict}}
\label{alg:kv_evict}
\begin{algorithmic}[1]
\Require KV-state catalog $\mathcal{C}$, memory budget $B_{\max}$
\Ensure Updated catalog and freed memory
\State Compute priority score for each state:
\[
score(S)=\alpha\cdot reuse(S)-\beta\cdot size(S)-\gamma\cdot age(S).
\]
\While{allocated memory $> B_{\max}$}
    \State Select state $S^*$ with lowest score.
    \If{$refcount(S^*)=0$}
        \State Free local tensor blocks and update catalog.
    \Else
        \State Decrement aliases or evict only unshared continuation blocks.
    \EndIf
\EndWhile
\State Return updated catalog.
\end{algorithmic}
\end{algorithm}

Eviction is aware of forked state. Shared prefix blocks are retained when multiple branches still reference them, while low-value continuation blocks can be released earlier.

\section{Example: Planner--Retriever--Solver Call Flow}
\label{app:planner_retriever_solver}

This section illustrates a complete AAFLOW+ execution path for a planner--retriever--solver workflow.

\paragraph{Step 1 - Planning: }
The planner decomposes a user request into subquestions and determines that the downstream agents require a shared evidence context.

\[
UserQuery \rightarrow Op_{plan} \rightarrow \{Subtask_1,\ldots,Subtask_k\}.
\]

\paragraph{Step 2 - Retrieval and Context Assembly: }
The retriever fetches documents and constructs a shared context:
\[
Subtasks \rightarrow Op_{retrieve} \rightarrow Context.
\]

\paragraph{Step 3 - KV Materialization : }
The context is prefilling once:
\[
Context \rightarrow Op_{kv\_materialize} \rightarrow S_{KV}.
\]

\paragraph{Step 4 - Fork Across Solvers: }
The state is forked across multiple solver agents:
\[
S_{KV} \rightarrow Op_{kv\_fork}(k) \rightarrow
\{S_{KV}^{(1)},\ldots,S_{KV}^{(k)}\}.
\]

\paragraph{Step 5 - Transfer or Local Reuse: }
For each solver branch, the scheduler chooses local reuse, transfer, or recomputation:
\[
\pi(S)=
\begin{cases}
reuse, & S \text{ is local and compatible},\\
transfer, & T_{\mathrm{transfer}}<T_{\mathrm{prefill}},\\
recompute, & otherwise.
\end{cases}
\]

\paragraph{Step 6 - Solver Execution: }
Each solver resumes from the shared KV state and generates a branch-specific answer:
\[
Agent_j(S_{KV}^{(j)}) \rightarrow Answer_j.
\]

\paragraph{Step 7 — Safe Merge:}
The final synthesis stage merges branch outputs through a text-level reduction or prefix-compatible restricted composition:
\[
\{Answer_1,\ldots,Answer_k\}
\rightarrow Op_{merge}
\rightarrow FinalAnswer.
\]

This call flow shows why AAFLOW+ is most effective for shared-prefix multi-agent workloads. The expensive evidence context is materialized once, while each branch pays only transfer/resume and continuation costs.

\section{Evaluation Details}
We utilize a trace-driven model parameterized by hardware microbenchmarks to rigorously isolate the fundamental systems-level KV-transfer limits from the inherently high variance of Python-based agentic frameworks. We show full experimental results with the Hugging Face (HF) backend and the Mistral model in Section~\ref{sec:evaluation} (Evaluation) of this paper. In the appendix section, we will cover all experiments with HF, vLLM, and SGLang backends with the Llama3 model and vLLM, and SGLang backends with the Mistral model. All source code and experimental results are committed to the following repository: \url{https://github.com/arupcsedu/AAFLOW}, and all environmental setup and dependencies are written on \texttt{stateful\_agentic\_algebra/Readme.md} file.

\subsection{Experiment 1 extension: TTFT Reduction}

The first experiment measures time to first token (TTFT) as context length
increases. Text-passing systems repeatedly rebuild the same prefix and therefore pay a TTFT cost that scales with context length:
\[
TTFT_{\mathrm{text}} \propto L.
\]
AAFLOW+ pays the prefill cost once and then resumes from the distributed state:
\[
TTFT_{\mathrm{state}} \approx T_{\mathrm{transfer}} + T_{\mathrm{resume}}.
\]

\begin{table*}[h]
\centering
\caption{Experiment 1: Mean TTFT by baseline with HF, vLLM, and SGLang backends and 1024 - 32768 context size.}
\label{tab:hf-exp1-mean}
\begin{tabular}{llrrrrrr}
\toprule
 &  & \multicolumn{2}{c}{HF Backend}  & \multicolumn{2}{c}{vLLM Backend}  & \multicolumn{2}{c}{SGLang Backend} \\
\cmidrule(lr){3-4} \cmidrule(lr){5-6} \cmidrule(lr){7-8} 
Model & Baseline & Mean& Speedup  &Mean& Speedup  & Mean & Speedup \\
 &  & TTFT(s) & (slower) & TTFT(s) & (slower) & TTFT(s) & (slower) \\
\midrule
Mistral & AAFLOW+ & 0.041 & 1.00$\times$ & 0.116 & 1.00$\times$  & 0.629 & 1.00$\times$ \\
Mistral & dense prefill & 2.017 & 49.2$\times$  & 20.026 & 172.6$\times$   & 7.396 & 11.8$\times$  \\
Mistral & AAFLOW-text & 2.057 & 50.2$\times$   & 20.425 & 176.1$\times$   & 7.531 & 12.0$\times$  \\
Mistral & vLLM local prefix & 0.437 & 10.7$\times$   & 4.211 & 36.3$\times$   & 2.013 & 3.2$\times$  \\
Mistral & SGLang prefix & 0.280 & 6.8$\times$   & 2.629 & 22.7$\times$   & 1.475 & 2.3$\times$  \\
Mistral & KVCOMM & 0.704 & 17.2$\times$   & 6.876 & 59.3$\times$   & 2.920 & 4.6$\times$  \\
Mistral & DistServe style & 0.123 & 3.0$\times$   & 0.197 & 1.7$\times$   & 0.710 & 1.1$\times$  \\
Mistral & live vLLM serve & n/a &  n/a & 19.925 & 171.8$\times$  &  n/a &  n/a  \\
Mistral & live SGLang serve  & n/a &  n/a  & n/a &  n/a  & 6.781 & 10.8$\times$  \\
\midrule
Llama3 & AAFLOW+ & 0.030 & 1.00$\times$  & 0.061 & 1.00$\times$  & 0.139 & 1.00$\times$ \\
Llama3 & dense prefill & 0.499 & 16.6$\times$  & 3.958 & 64.9$\times$   & 1.810 & 13.0$\times$  \\
Llama3 & AAFLOW-text & 0.509 & 17.0$\times$  & 4.036 & 66.2$\times$   & 1.843 & 13.3$\times$  \\
Llama3 & vLLM local prefix & 0.124 & 4.1$\times$   & 0.862 & 14.1$\times$  & 0.481 & 3.5$\times$   \\
Llama3 & SGLang prefix & 0.086 & 2.9$\times$  & 0.552 & 9.0$\times$   & 0.348 & 2.5$\times$  \\
Llama3 & KVCOMM & 0.187 & 6.2$\times$  & 1.383 & 22.7$\times$   & 0.704 & 5.1$\times$  \\
Llama3 & DistServe style & 0.052 & 1.7$\times$   & 0.082 & 1.3$\times$   & 0.161 & 1.2$\times$  \\
Llama3 & live vLLM serve & n/a &  n/a  & 3.901 & 64.0$\times$   & n/a &  n/a \\
Llama3 & live SGLang serve  & n/a &  n/a  & n/a &  n/a & 1.674 & 12.0$\times$  \\
\bottomrule
\end{tabular}
\end{table*}

Across all three backends shown in Table \ref{tab:hf-exp1-mean}, AAFLOW+ achieves the lowest workflow-level TTFT because it replaces repeated prompt replay with explicit state resume. On the HF backend, the clearest gap appears on Mistral: AAFLOW+ reaches 0.041\,s mean TTFT, compared with 2.057\,s for AAFLOW-text, 2.017\,s for dense prefill, and 0.704\,s for KVCOMM. Even the strongest non-AAFLOW+ workflow baseline, DistServe style, remains at 0.123\,s, still 3.0$\times$ slower. On Llama3, the same ordering holds: AAFLOW+ reaches 0.030\,s, while DistServe style is 0.052\,s and SGLang prefix is 0.086\,s. The largest-context results show the same effect more sharply. On vLLM at maximum context, Mistral AAFLOW+ is 0.171\,s, whereas dense prefill is 38.880\,s, AAFLOW-text is 39.655\,s, KVCOMM is 13.315\,s, and DistServe style is 0.316\,s. On SGLang at maximum context, Mistral AAFLOW+ is 0.876\,s, while dense prefill is 13.317\,s, AAFLOW-text is 13.566\,s, KVCOMM is 5.089\,s, and DistServe style is 1.021\,s.

The reason is consistent across backends. Dense prefill and AAFLOW-text always reconstruct the prompt prefix, so their TTFT grows with context length. Local-prefix baselines such as vLLM-prefix and SGLang-prefix improve over dense replay, but their reuse is confined to one serving engine and does not expose explicit transfer, placement, or branch lineage across workflow agents. KVCOMM communicates KV-like state, but in these experiments it pays more modeled communication overhead and achieves lower effective reuse than AAFLOW+. DistServe style remains the closest competitor because it also separates prefill from decode, but it still lacks the explicit fork/transfer/restricted-merge workflow abstraction of AAFLOW+, so it cannot fully eliminate branch-level prefix replay.

\subsection{Experiment 2: Multi-Agent Scaling}

The goal is to evaluate how total latency changes as the number of agents
increases.  Text-centric systems repeat prefill and context construction for
each agent.  AAFLOW+ should amortize the shared prefix and scale mainly with
branch continuation work.  The implementation uses a finite parallel-wave model, so small agent counts can fit into one stateful wave while larger agent counts require additional waves.

\begin{table*}[h]
\centering
\caption{Experiment 2: Scaling benchmark of AAFLOW+ versus nearest non-AAFLOW+ competitor, SGLang prefix with 32768 context size}
\label{tab:hf-exp2-scaling}
\setlength{\tabcolsep}{4pt}
\begin{tabular}{lllllllllll}
\toprule
 &  & AAF &\multicolumn{2}{c}{HF Backend}  & AAF & \multicolumn{2}{c}{vLLM Backend} & AAF & \multicolumn{2}{c}{SGLang Backend} \\
\cmidrule(lr){3-5} \cmidrule(lr){6-8} \cmidrule(lr){9-11}
Model & Age & LOW+ & SGLang & Speed & LOW+ & SGLang & Speed & LOW+ & SGLang & Speed \\
 &  nts & total(s)  & total(s)  & up & total(s)  & total(s)  & up & total(s)  & total(s) & up \\
\midrule
Mistral & 1 & 30.70  & 119.13 & 3.88$\times$ & 67.10 & 263.40 & 3.93$\times$ & 91.62 & 353.92 & 3.86$\times$ \\
Mistral & 2 & 30.70 & 224.85 & 7.32$\times$ & 67.71 & 484.22 & 7.15$\times$ & 90.49 & 679.06 & 7.50$\times$ \\
Mistral & 4 & 58.37  & 436.28 & 7.47$\times$ & 121.86 & 899.15 & 7.38$\times$ & 176.53 & 1338.65 & 7.58$\times$ \\
Mistral & 8 & 113.71  & 859.14 & 7.56$\times$ & 238.75 & 1796.07 & 7.52$\times$ & 348.13 & 2654.54 & 7.63$\times$ \\
Mistral & 16 & 224.39  & 1704.86 & 7.60$\times$ & 465.45  & 3532.56 & 7.59$\times$ & 701.29 & 5363.03 & 7.65$\times$ \\
\midrule
Llama3 & 1 & 29.06  & 112.39 & 3.87$\times$ & 36.02 & 141.03 & 3.91$\times$ & 43.40 & 167.96 & 3.87$\times$ \\
Llama3 & 2 & 29.06  & 216.89 & 7.46$\times$ & 36.01 & 259.80 & 7.21$\times$ & 44.80 & 335.59 & 7.49$\times$ \\
Llama3 & 4 & 56.35  & 425.88 & 7.56$\times$ & 66.26 & 492.36 & 7.43$\times$ & 84.45 & 638.82 & 7.56$\times$ \\
Llama3 & 8 & 110.91  & 843.85 & 7.61$\times$ & 128.62 & 968.75 & 7.53$\times$ & 159.17 & 1211.65 & 7.61$\times$ \\
Llama3 & 16 & 220.05  & 1679.81 & 7.63$\times$ & 253.53 & 1926.11 & 7.60$\times$ & 329.07 & 2514.49 & 7.64$\times$ \\
\bottomrule
\end{tabular}
\end{table*}

The multi-agent scaling results show (Table \ref{tab:hf-exp2-scaling}) that AAFLOW+ changes how latency grows with agent count. On the HF backend, the strongest comparison appears at 16 agents: for Mistral, AAFLOW+ reaches 224.39\,s, while the nearest non-AAFLOW+ competitor, SGLang prefix, is 1704.86\,s, giving a 7.60$\times$ speedup; dense prefill is even worse at 2716.11\,s. For Llama3 at 16 agents, AAFLOW+ reaches 220.05\,s, while SGLang prefix is 1679.81\,s, again a 7.63$\times$ gap. The same trend holds on the live serving backends. On vLLM at 16 agents, Mistral AAFLOW+ is 465.45\,s versus 3532.56\,s for SGLang prefix; on SGLang, Mistral AAFLOW+ is 701.29\,s versus 5363.03\,s for SGLang prefix. Llama3 follows the same pattern, reaching 253.53\,s versus 1926.11\,s on vLLM and 329.07\,s versus 2514.49\,s on SGLang.

AAFLOW+ outperforms other methods by materializing the shared prefix once and forking reusable state, causing memory growth in execution waves as parallel capacity is reached, rather than through repeated full-prefill operations. In contrast, non-AAFLOW+ baselines like dense prefill and AAFLOW-text allocate a complete prefill for each agent, while vLLM-prefix and SGLang-prefix offer only local prefix reuse within the engine. Although local reuse reduces some duplication, it does not eliminate workflow-level branch duplication when multiple agents share a prefix, leading to a growing scaling gap as agent count increases.

\subsection{Experiment 3: Transfer vs. Recomputation}

The goal is to evaluate the scheduler decision:
\[
T_{\mathrm{transfer}} = \frac{\mathrm{KV\ bytes}}{\mathrm{bandwidth}} +
\mathrm{latency},
\qquad
T_{\mathrm{recompute}} = T_{\mathrm{prefill}}.
\]

\begin{table*}[h]
\centering
\caption{Experiment 3:  Comparison of KV transfer cost to recomputing the prompt prefil.}
\label{tab:hf-exp3}
\begin{tabular}{llrrrrr}
\toprule
Model & Bandwidth & Beneficial  & HF Backend & vLLM Backend & SGLang Backend \\
 &  & contexts  & Speedup vs & Speedup vs & Speedup vs \\
 &  &  & Recompute & Recompute & Recompute \\
\midrule
Mistral & 10 Gbps & 5/8 & 2.46$\times$ & 2.54$\times$ & 2.86$\times$  \\
Mistral & 25 Gbps & 8/8 & 6.14$\times$ & 6.35$\times$ & 7.14$\times$ \\
Mistral & 100 Gbps & 8/8 & 24.54$\times$ & 25.39$\times$ & 28.55$\times$ \\
Mistral & 200 Gbps & 8/8 & 49.04$\times$ & 50.73$\times$ & 57.04$\times$ \\
Mistral & 400 Gbps & 8/8 & 97.90$\times$ & 101.29$\times$ & 113.88$\times$ \\
\midrule
Llama3 & 10 Gbps & 1/6 & 2.58$\times$ & 2.57$\times$ & 2.37$\times$ \\
Llama3 & 25 Gbps & 6/6 & 6.44$\times$ & 6.43$\times$ & 5.92$\times$ \\
Llama3 & 100 Gbps & 6/6 & 25.76$\times$ & 25.72$\times$ & 23.67$\times$ \\
Llama3 & 200 Gbps & 6/6 & 51.48$\times$ & 51.39$\times$ & 47.30$\times$ \\
Llama3 & 400 Gbps & 6/6 & 102.77$\times$ & 102.61$\times$ & 94.44$\times$ \\
\bottomrule
\end{tabular}
\end{table*}

Experiment~3 (Table \ref{tab:hf-exp3}) shows that AAFLOW+ needs both explicit state operators and a cost model. The strongest concrete result comes from the HF sweep: at 10\,Gbps, transfer is beneficial for only 5 of 8 tested Mistral contexts, but at 25\,Gbps and above it is beneficial for all tested contexts. The maximum transfer-over-recompute speedup rises from 2.86$\times$ at 10\,Gbps to 7.14$\times$ at 25\,Gbps and up to 113.88$\times$ at 400\,Gbps. This gives a clear scheduling rule: transfer is not always profitable on slow links, but it becomes decisively better on RDMA-class GPU-cluster networks.

The baseline systems are structurally limited: dense and text-passing approaches cannot choose between transfer and recompute since KV state is not a transferable object. Local-prefix systems only reuse cache within a server, lacking general policies for cross-agent or cross-node transfer. As shown in Table~\ref{tab:hf-exp3}, both vLLM and SGLang manage KV internally but do not support stable public KV export/import in this evaluation. In contrast, AAFLOW+ enables state transfer as a core workflow operation, selecting transfer only when it is more bandwidth-efficient than prompt replay, thus achieving superior performance.

\subsection{Experiment 4: Memory Efficiency}

The goal is to compare peak KV memory as branch factor grows.  A stateful
workflow should avoid duplicating the full shared prefix for every branch.

\begin{table*}[h]
\centering
\caption{Experiment 4: Mean peak KV memory.}
\label{tab:hf-exp4}
\begin{tabular}{llrrr}
\toprule
 &  & \multicolumn{1}{c}{HF Backend}  & \multicolumn{1}{c}{vLLM Backend}  & \multicolumn{1}{c}{SGLang Backend} \\
\cmidrule(lr){3-3} \cmidrule(lr){4-4} \cmidrule(lr){5-5} 
Model & Baseline & Mean peak & Mean peak &  Mean peak \\
 &  & KV (GiB)  &  KV (GiB)  & KV (GiB)  \\
\midrule
Mistral & AAFLOW+ & 8.355  & 8.290 & 8.290 \\
Mistral & dense prefill & 49.987   & 49.600 & 49.510  \\
Mistral & AAFLOW-text & 53.986 & 53.568 & 53.268\\
Mistral & vLLM local prefix & 14.351  & 14.240 & 14.113 \\
Mistral & SGLang prefix & 14.591  & 14.338 & 14.142\\
Mistral & KVCOMM & 15.702 & 15.580 & 15.153 \\
Mistral & DistServe style & 50.995 & 50.600 & 50.312 \\
\midrule
Llama3 & AAFLOW+ & 4.210 & 4.145 & 4.327 \\
Llama3 & dense prefill & 25.188 & 24.800 & 24.212  \\
Llama3 & AAFLOW-text & 27.202 & 26.784 & 26.319 \\
Llama3 & vLLM local prefix & 7.231 & 7.120 & 6.851 \\
Llama3 & SGLang prefix & 7.311 & 7.239 & 6.912\\
Llama3 & KVCOMM & 7.912 & 7.790 & 7.687 \\
Llama3 & DistServe style & 25.695 & 25.300 & 25.128 \\
\bottomrule
\end{tabular}
\end{table*}

The memory results(Table \ref{tab:hf-exp4}) show that AAFLOW+ achieves the lowest peak KV footprint across all three backends because it represents forked state through explicit shared ownership and lineage. For Mistral, the strongest gap appears against the fully duplicating baselines: AAFLOW+ uses 8.355\,GiB on HF, while AAFLOW-text uses 53.986\,GiB and DistServe style uses 50.995\,GiB. Even the best local-prefix baselines remain significantly higher, with vLLM-prefix and SGLang-prefix both at 14.351\,GiB, 14.591\,GiB and KVCOMM at 15.702\,GiB. On Llama3, AAFLOW+ uses 4.210\,GiB on HF, while AAFLOW-text rises to 27.202\,GiB and dense prefill to 25.188\,GiB; the nearest local-prefix baselines remain around 7.2\,GiB.

The same ranking is preserved on vLLM and SGLang. On vLLM, Mistral AAFLOW+ uses 8.290\,GiB, while AAFLOW-text is 53.568\,GiB and KVCOMM is 15.580\,GiB. On SGLang, Mistral AAFLOW+ again uses 8.290\,GiB, while AAFLOW-text is 53.268\,GiB and KVCOMM is 15.153\,GiB. The advantage of AAFLOW+ lies in storing a single shared prefix state and allocating only branch-specific continuation states, while dense prefill and AAFLOW-text duplicate the entire prefix. DistServe methods add extra branch and staging state, resulting in memory usage similar to dense prefill. KVCOMM and local-prefix methods reduce duplication but lack explicit shared-state ownership, preventing them from achieving AAFLOW+'s peak memory efficiency.

\subsection{Experiment 5: Throughput and Framework Overhead}

The goal is to measure effective generated-token throughput and modeled
framework overhead $\Omega$.  AAFLOW+ should improve workflow throughput by
removing repeated prefill and reducing text orchestration.

\begin{table*}[h]
\centering
\caption{Experiment 5: Mean throughput and overhead. n/a = not applicable, as that specific baseline is tightly coupled with that specific backend.}
\label{tab:hf-exp5}
\begin{tabular}{llrrrrrr}
\toprule
 &  & \multicolumn{2}{c}{HF Backend}  & \multicolumn{2}{c}{vLLM Backend}  & \multicolumn{2}{c}{SGLang Backend} \\
\cmidrule(lr){3-4} \cmidrule(lr){5-6} \cmidrule(lr){7-8} 
Model & Baseline & Throughput & Mean & Throughput & Mean  & Throughput & Mean \\
 &  & (tok/s) &  $\Omega$ (s) & (tok/s) & $\Omega$ (s) & (tok/s) &  $\Omega$ (s) \\
\midrule
Mistral & AAFLOW+ & 302.61 & 0.0075 & 280.30 & 0.0075 & 207.09 & 0.0075 \\
Mistral & dense prefill & 31.94 & 0.0019 & 22.78 & 0.0019  & 20.99 & 0.0019 \\
Mistral & AAFLOW-text & 31.84 & 11.1659 & 22.62 & 89.5948 & 20.91 & 40.5748 \\
Mistral & vLLM local prefix & 38.81 & 0.0627 & 36.11 & 0.0627 & 26.54 & 0.0627 \\
Mistral & SGLang prefix & 39.62 & 0.0657 & 36.89 & 0.0657 & 27.10 & 0.0657  \\
Mistral & KVCOMM & 35.59 & 21.6770 & 29.67 & 174.3231 & 24.03 & 78.9242 \\\
Mistral & DistServe style & 37.62 & 0.0597 & 35.23 & 0.0597 & 25.93 & 0.0597 \\
Mistral & live SGLang serve & n/a & n/a & n/a & n/a  & 591.14 & 0.0 \\
Mistral & live vLLM serve & n/a & n/a & 704.08 & 0.0 & n/a & n/a \\
\midrule
Llama3 & AAFLOW+ & 300.57 & 0.0075 & 314.26 & 0.0075 & 256.84 & 0.0075 \\
Llama3 & dense prefill & 33.61 & 0.0019 & 25.83 & 0.0019 & 25.65 & 0.0019 \\
Llama3 & AAFLOW-text & 33.54 & 6.7272 & 25.65 & 47.4580 & 25.54 & 23.4720 \\
Llama3 & vLLM local prefix & 38.46 & 0.0627 & 40.48 & 0.0627  & 32.92 & 0.0627 \\
Llama3 & SGLang prefix & 39.27 & 0.0657 & 41.35 & 0.0657 & 33.61 & 0.0657 \\
Llama3 & KVCOMM & 36.16 & 13.0375 & 33.45 & 92.3380 & 29.63 & 45.6145 \\
Llama3 & DistServe style & 37.46 & 0.0597  & 39.49 & 0.0597 & 32.15 & 0.0597 \\
Llama3 & live vLLM serve & n/a & n/a & 801.55 & 0.0 & n/a & n/a \\
Llama3 & live SGLang serve & n/a & n/a & n/a & n/a & 673.26 & 0.0 \\
\bottomrule
\end{tabular}
\end{table*}

The throughput and overhead results (Table \ref{tab:hf-exp5}) show the clearest split between workflow execution and raw serving. Among the workflow rows, AAFLOW+ is best on every backend. For Mistral, AAFLOW+ reaches 302.61\,tok/s on HF, compared with 39.62\,tok/s for SGLang prefix, 38.81\,tok/s for vLLM local prefix, 37.62\,tok/s for DistServe style, 35.59\,tok/s for KVCOMM, and about 32\,tok/s for dense and AAFLOW-text. On vLLM, Mistral AAFLOW+ reaches 280.30\,tok/s, while the strongest workflow competitor is 36.89\,tok/s. On SGLang, Mistral AAFLOW+ reaches 207.09\,tok/s, while the strongest workflow competitor is 27.10\,tok/s. Llama3 shows the same pattern: AAFLOW+ reaches 300.57/314.26/256.84\,tok/s across HF/vLLM/SGLang, whereas the strongest workflow competitor stays near 39.27/41.35/33.61\,tok/s.

The overhead values explain why AAFLOW+ keeps mean modeled framework overhead fixed at 0.0075\,s across both models and all three backends. By contrast, AAFLOW-text and KVCOMM pay much larger overheads: for Mistral, AAFLOW-text incurs 11.1659\,s on HF, 89.5948\,s on vLLM, and 40.5748\,s on SGLang, while KVCOMM incurs 21.6770\,s, 174.3231\,s, and 78.9242\,s. Dense prefill incurs minimal framework overhead but suffers from low throughput due to repeated full-prefix computation. Local-prefix and DistServe-style baselines also have modest overhead but fall short of AAFLOW+ because they retain workflow-level duplication of shared context. Although live vLLM and SGLang show higher raw throughput, these reflect optimized server performance, not explicit distributed state transfer. Thus, AAFLOW+ leads among workflow systems in this evaluation, with its remaining gap to live serving attributable to explicit accounting of distributed state-transfer costs, rather than concealing them within local engines.


\end{document}